\date{\today}

\documentclass[aps,prb,twocolumn,superscriptaddress,groupedaddress]{revtex4}  % for review and submission
\usepackage{dcolumn,graphicx,amsmath,amssymb,algorithm,algpseudocode}
\usepackage{todonotes}
\usepackage{qcircuit}

\newcommand{\pdiff}[2]{\frac{\partial #1}{\partial #2}}

\begin{document}

\definecolor{brickred}{rgb}{.72,0,0} 

\title{
A Jacobi Diagonalization and Anderson Acceleration Algorithm For Variational
Quantum Algorithm Parameter Optimization 
}

\author{Robert M. Parrish}
\email{rob.parrish@qcware.com}
\affiliation{
QC Ware Corporation, Palo Alto, CA 94301
}

\author{Joseph T. Iosue}
\affiliation{
QC Ware Corporation, Palo Alto, CA 94301
}

\author{Asier Ozaeta}
\affiliation{
QC Ware Corporation, Palo Alto, CA 94301
}

\author{Peter L. McMahon}
\affiliation{
E.\,L.~Ginzton Laboratory, Stanford University, Stanford, CA 94305
    }
\affiliation{
QC Ware Corporation, Palo Alto, CA 94301
}

\begin{abstract} 
The optimization of circuit parameters of variational quantum algorithms such as
the variational quantum eigensolver (VQE) or the quantum approximate
optimization algorithm (QAOA) is a key challenge for the practical deployment of
near-term quantum computing algorithms. Here, we develop a hybrid
quantum/classical optimization procedure inspired by the Jacobi diagonalization
algorithm for classical eigendecomposition, and combined with Anderson acceleration.
In the first stage, analytical tomography fittings are performed for a local
cluster of circuit parameters via sampling of the observable objective function
at quadrature points in the circuit angles. Classical optimization is used to
determine the optimal circuit parameters within the cluster, with the other
circuit parameters frozen. Different clusters of circuit parameters are then
optimized in ``sweeps,'' leading to a monotonically-convergent fixed-point
procedure.  In the second stage, the iterative history of the fixed-point Jacobi
procedure is used to accelerate the convergence by applying Anderson
acceleration/Pulay's direct inversion of the iterative subspace (DIIS). This
Jacobi+Anderson method is numerically tested using a quantum circuit simulator
(without noise) for a representative test case from the multistate, contracted
variant of the variational quantum eigensolver (MC-VQE), and is found to be
competitive with and often faster than Powell's method and L-BFGS. 
\end{abstract}

% 03.67.Ac	Quantum algorithms, protocols, and simulations
% 31.10.+z	Theory of electronic structure, electronic transitions, and chemical binding
% 31.15.-p	Calculations and mathematical techniques in atomic and molecular physics 
% 03.65.-w	Quantum mechanics
% \pacs{03.67.Ac,31.10.+z,31.15.-p}

\maketitle

\section{Introduction}

The past few years have witnessed the arrival of a wave of hybrid
quantum/classical algorithms\cite{ 
Peruzzo:2014:4213,
Farhi:2014:Quantum}
with compelling properties for use
on noisy intermediate-scale quantum (NISQ) devices.\cite{Preskill:2018:79} A
universal feature of these methods is that they involve the hybrid
quantum/classical optimization of a real quantum observable expectation value
$\mathcal{O}$ with respect to a polynomial number of quantum circuit parameters
$\{ \theta_g \}$. For a given set of circuit parameters, the quantum observable
expectation value can be determined by statistically sampling the output
measurements of the quantum circuit, 
\begin{equation}
\label{eq:O_quantum}
\mathcal{O} (\{ \theta_g \})
=
\langle \vec 0 | 
\hat U^\dagger (\{ \theta_g \})
\hat O
\hat U (\{ \theta_g \})
| \vec 0 \rangle
\end{equation}
\begin{equation}
\label{eq:O_dm}
=
\sum_{k}
\gamma_{k} (\{ \theta_g \})
h_{k}
\end{equation}
Here $| \vec 0 \rangle$ is the all-zeros starting qubit state for $N$ qubits,
$\hat U (\{ \theta_g \})$ is the $2^N$-dimensional Hilbert space unitary
operator parameterized by the quantum circuit, and $\hat O$ is a Hermitian
operator defining the desired quantum observable expectation value. Generally
$\hat O$ is defined in a problem-specific way in terms of a polynomial number of
Pauli strings $\{ \hat P_{k} \}$ accompanied by problem-specific matrix elements
$\{ h_{k} \}$ (determined classically before the quantum algorithm is applied),
\begin{equation}
\label{eq:O}
\hat O
\equiv
\sum_{k}
h_{k} 
\hat P_{k}
\end{equation}
In practice, statistical observation of the quantum circuit is used to
determine the density matrix elements,
\begin{equation}
\label{eq:dm}
\gamma_{k}
\equiv
\langle \vec 0 | 
\hat U^\dagger (\{ \theta_g \})
\hat P_{k}
\hat U (\{ \theta_g \})
| \vec 0 \rangle
\end{equation}
These are subsequently contracted classically with the matrix elements $\{ h_k
\}$ as in Equation \ref{eq:O_dm}. At this point, we have a recipe (which
involves repeatedly observing the output of a quantum circuit) for the evaluation of the
observable expectation value
$\mathcal{O}$ at a given set of circuit parameters $\{ \theta_g \}$. The
remaining task, and the major driver for this study, is to classically optimize
the $N_g$-dimensional objective function $\mathcal{O} ( \{ \theta_g \})$ with as
few observable expectation value evaluations as possible, and in the presence of
both statistical and physical-device-imperfection noise channels. 

Quantum methods of this class include the variational quantum eigensolver\cite{ 
Peruzzo:2014:4213,
McClean:2016:023023,
OMalley:2016:031007,
Kandala:2017:242,
McClean:2017:X,
Romero:2018:104008}
(VQE) and the quantum approximate optimization
algorithm\cite{Farhi:2014:Quantum} (QAOA). VQE was initially developed for
quantum chemistry applications such as the approximation of the ground-state
energy of the electronic wavefunction of small molecular complexes.  VQE has
since seen widespread extension of application to the simulation of electronic,
magnetic, and excitonic Hamiltonians.  Extensions to VQE such as the folded
spectrum,\cite{Peruzzo:2014:4213}
orthogonality-constrained\cite{Higgott:2018:X,Lee:2018:JCTC} (OC-VQE), quantum
subspace expansion\cite{McClean:2017:042308,Colless:2018:011021} (QSE-VQE), the
subspace search\cite{Nakanishi:2018:VQE} (SS-VQE), and  multistate, contracted
\cite{Parrish:2019:MCVQE} (MC-VQE) extensions have enabled the computation of
excited
states\cite{Higgott:2018:X,Lee:2018:JCTC,McClean:2017:042308,Colless:2018:011021,Nakanishi:2018:VQE,Parrish:2019:MCVQE}
and transition properties.\cite{Parrish:2019:MCVQE} A recent advance has even
enabled a post-quantum classical correction from perturbation theory to enhance
the accuracy of the method.\cite{Takeshia:2019:PT2} VQE and its extensions have
been deployed on physical quantum hardware to simulate the ground and
excited-state energies of small molecules and magnetic systems, including
deployment on
superconducting\cite{Peruzzo:2014:4213,Kandala:2017:242,Colless:2018:011021}
and ion-trap\cite{Nam:2019:IonWater} quantum hardware.

The quantum approximate optimization algorithm\cite{Farhi:2014:Quantum} (QAOA) is another prominent variational quantum
algorithm; it is designed to try solve classical combinatorial-optimization
problems over binary variables, such as MAX-CUT or classical Ising optimization.
QAOA uses alternating unitary evolutions of a cost-function Hamiltonian and a
driver Hamiltonian, and canonically each evolution unitary has associated with
it a real-valued parameter. It is these parameters that are varied in an
attempt to have the circuit return optimal choices for the binary variables.
QAOA has broad potential application due partially to the fact that many
NP-hard optimization problems can be mapped to Ising form with low overhead,\cite{Lucas:2014:Ising,  glover2018tutorial} or variations of QAOA may more
directly allow solution of constrained binary optimization problems.\cite{hadfield2019quantum} Much of the recent body of work exploring use cases
for quantum annealing is also directly relevant to QAOA, since it focuses on
problems that can be mapped to Ising form.\cite{smelyanskiy2012near,biswas2017nasa,orus2019quantum} While some initial
implementations of QAOA used off-the-shelf gradient-free classical optimization
methods,\cite{farhi2017quantum,otterbach2017unsupervised} a substantial body
of work has developed exploring off-the-shelf and customized use of
gradient-free and gradient-based methods.\cite{moll2018quantum,gilyen2019optimizing,crooks2018performance,zhou2018quantum,brandao2018fixed}
There has also been recent work that systematically compares six off-the-shelf
optimization methods for QAOA.\cite{nannicini2019performance}

Methods for optimizing variational circuits using analytical gradients have also
seen rapid development over the past two years\cite{Guerreschi:2017:Opt,Li:2017:Hybrid,Mitarai:2018:Quantum,Verdon:2018:Universal,Bergholm:2018:Pennylane,Schuld:2019:Evaluating}.
These techniques have been applied to examples in QAOA, VQE, as well as machine
learning (in particular, quantum neural networks). It has also recently been
noted that variational methods on random quantum circuits may suffer from
``barren plateaus'' filled with vanishing gradients;\cite{McClean:2018:4812}
this needs to be taken into consideration when using gradient-based methods.

\begin{figure*}[h!]
\begin{center}
\includegraphics[width=6.4in]{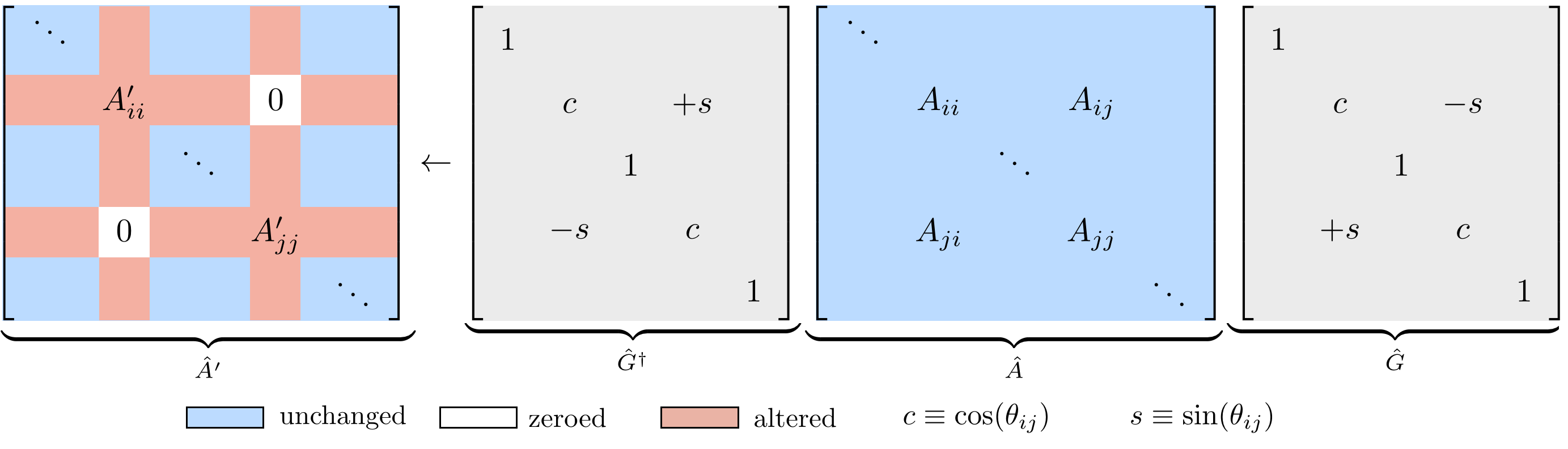}
\caption{Schematic of classical Jacobi diagonalization algorithm local move for
pivot index $(i,j)$. A local $2\times 2$ eigenproblem is solved at the $(i,j)$
pivot index, determining a Givens rotation $\hat G$ which explicitly zeros
$A_{ij}'$.  This alters the $i$-th and $j$-th rows and columns of the matrix
$\hat A'$.  Subsequently, the process is repeated in ``sweeps'' over different
sets of $(i,j)$ pivot indices. Each step reduces the off-diagonal weight of
matrix $\hat A$, so the convergence is monotonic in an objective function
defined as $O (\{ \theta_{ij} \}) = + \sum_{i\neq j} A_{ij}^2$ or $O (\{
\theta_{ij} \}) = - \sum_{i} A_{ii}^2$.}
\label{fig:jacobi}
\end{center}
\end{figure*}

The inspiration for the methods we develop in this work is a much older class of
optimization/root-finding
algorithms: classical fixed point methods. One key method within this class is
the Jacobi algorithm for the classical diagonalization of symmetric matrices,\cite{Jacobi:1846:51}
depicted in Figure \ref{fig:jacobi}. The Jacobi diagonalization fixed-point
iteration works by successively identifying an $(i,j)$ ``pivot index'' within
matrix $\hat A$, and then applying an orthogonal transformation by a Givens
matrix to zero the off-diagonal element $A_{ij}$ -- i.e., exactly solving a local
$2\times 2$ subproblem. This is equivalent to diagonalizing the $2\times2$
submatrix of $\hat A$ corresponding to $A_{ii}$, $A_{ij}$ and $A_{jj}$, for
which the angle of the Givens matrix can be determined analytically. In the
process, the other elements of the $i$-th and $j$-th rows and columns of $\hat
A$ are mixed. This local optimization can then be repeated across different
sequences of $(i,j)$ pivot indices, hopefully making progress toward global
diagonalization of $\hat A$. Of note, each discrete $(i,j)$ optimization move is
negative definite, so the overall algorithm converges monotonically. Many
possible choices for the selection of the sequence of $(i,j)$ pivots exist:
Searching the matrix at each step for the largest off-diagonal element
$|A_{ij}|$ to determine the $(i,j)$ pivot yields the fastest convergence rates,
but the search itself can be prohibitively expensive. Randomized $(i,j)$
selection is often used, though it introduces nondeterminism into the
fixed-point procedure (which will make forthcoming convergence acceleration
procedures difficult to apply).  Alternatively, simply ``sweeping'' over all of
the $(i > j)$ pairs in a predetermined order often provides acceptable
convergence within a simple and deterministic recipe. In practice, the Jacobi
diagonalization algorithm converges rapidly for diagonally dominant matrices,
and has compelling features for classical parallelization, though QR-based
methods have generally superseded Jacobi methods for dense diagonalization on
classical computing hardware.\cite{Karp:1987:0167} The Jacobi diagonalization
algorithm has been extended to Hermitian and non-Hermitian matrices and to to
singular value decomposition.\cite{Golub:2000:35} The Jacobi diagonalization
algorithm is also widely used in the approximate simultaneous diagonalization
of multiple matrices,\cite{Bunse:1993:927,Cardoso:1996:161} e.g., in domain
applications in orbital
localization,\cite{Boys:1960:296,Edmiston:1963:457,Pipek:1989,Knizia:2013:4834}
and discrete variable representation grid
determination,\cite{Dawes:2004:726,Dawes:2005:134101,Dawes:2006:054102,Parrish:2013:194107}
in computational chemistry.

The obvious disadvantage of local-move fixed-point algorithms like the Jacobi
diagonalization algorithm is that while each local move is optimal for the
given subproblem, the aggregate set of local moves are not necessarily
cooperative with respect to the global problem. This can lead to slow though
definitionally monotonic convergence. For instance, two or more different pivots
may ``slosh'' off-diagonal weight back and forth in a way that does not promote
rapid global convergence. To combat this, one can imagine extending the Jacobi
diagonalization algorithm to pairs, clusters, or even large subblocks of the
overall matrix. In practice, this often promotes more-rapid global convergence
at the cost of much more involved computations for the solution of each local
pair, cluster, or subblock solution. Moreover, the selection of the sets of
pivot clusters generally requires heuristics or special knowledge of the
structure of the problem. We will pursue a cluster approach below, but note that
one more piece is needed to complete the recipe: some procedure which develops
an effective or approximate global picture of the couplings between all pivot or
cluster moves within a Jacobi sweep, and which promotes global cooperation
between these moves. 

Such a procedure is the fixed-point algorithm doppelg\"anger of Krylov subspace
methods for iterative linear algebra problems or low-rank Hessian updating
approaches in quasi-Newton approximation methods, but here the application is to
sequence acceleration of a fixed-point-iteration method such as complete Jacobi
sweeps. Numerous examples of such fixed-point acceleration methods exist in the
mathematical physics literature, for example, Richardson extrapolation, Aitken's
delta-squared process, successive overrelaxation, and damping. One particularly
compelling method is the class of fixed-point sequence acceleration methods
developed by Anderson in 1965.\cite{Anderson:1965:547} Anderson acceleration has
been applied in many domain problems in numerical science and
engineering.\cite{walker2011anderson} Within the chemical physics literature, a
highly similar acceleration method was independently developed by Pulay in
1980,\cite{Pulay:1980:393} and extended in 1982,\cite{Pulay:1982:556} and is
known as direct inversion of the iterative subspace (DIIS). DIIS is widely used
to stabilize and accelerate the myriad classes of nonlinear equations appearing
in electronic structure theory, notably the self-consistent field (SCF)
equations,\cite{Pulay:1980:393,Pulay:1982:556,hamilton1986direct,hutter1994electronic}
coupled cluster (CC) equations,\cite{scuseria1986accelerating}
coupled-perturbed response equations, and even as an alternative to
quasi-Newton methods for the optimization of molecular
geometries.\cite{csaszar1984geometry} All Anderson/Pulay DIIS sequence
acceleration methods work by examining the iterative history of a sequence of
state vectors (e.g., the circuit angles $\{ \theta_g \}^{k}$ at each iteration
$k$) and a corresponding sequence of approximate error vectors $\{ \tilde
\epsilon_{g} \}^{k}$. An improved/extrapolated state vector of the form $\{
\theta_{g} \}^{\prime k} \equiv \sum_{i}^{k} c_i \{ \theta_g \}^{k}$ is
proposed, with the coefficients $c_i$ chosen to minimize the sum of squares of
the improved approximate error vector $\{ \epsilon_{g} \}^{\prime k} \equiv
\sum_{i}^{k} c_i \{ \epsilon_g \}^{k}$, subject to the normalization constraint
$\sum_{i}^{k} c_i = 1$. Many variations of Anderson and Pulay DIIS methods have
been developed\cite{kudin2002black,chen2011listb,hu2017projected} with
different choices of approximate error vectors and placements within
fixed-point iterative procedures.

In the present work, we develop a new local fixed-point-iteration plus global
sequence acceleration optimization algorithm for general variational quantum
circuit algorithms. First, we develop analytical formulae for the tomography of
the observable expectation values as a function of a cluster of $M$ circuit
parameters. Together with a $3^M$-point Fourier quadrature for the determination
of the tomography parameters, this allows us to obtain partial tomography for
each cluster of circuit parameters with a handful of observable evaluations at
widely spaced angles. Next, we classically optimize the objective function
within the tomography function for each cluster of angles. Then we sweep over
different clusters of angles in a Jacobi-diagonalization-like fixed point
procedure. Several sweep recipes are explored, such as all single-angle sweeps
(Jacobi-1), all double-angle sweeps (Jacobi-2), and locally selected clusters
based on quantum circuit structure (Jacobi-Gen), plus randomized-order
extensions (e.g., Jacobi-1-Rand). We also augment the procedure
with Anderson-type global sequence acceleration, using either the traditional
Anderson-type recipe with the fixed-point move length as the error vector
(Jacobi-$F$-Anderson) or the SCF Pulay DIIS-type recipe with the objective
function gradient as the error vector (Jacobi-$F$-Pulay).  We test the developed
methods against Powell's gradient-free conjugate direction
method\cite{Powell:1964:155} and the L-BFGS gradient-based quasi-Newton
method\cite{byrd1995limited} for MC-VQE+AIEM\cite{Parrish:2019:MCVQE} circuit
parameter optimization for an $N=4$ subset of bacteriochlorophylls in the B850
ring of LHII.

\section{Method Development}

\begin{figure*}[h!]
\begin{center}
\includegraphics[width=6.5in]{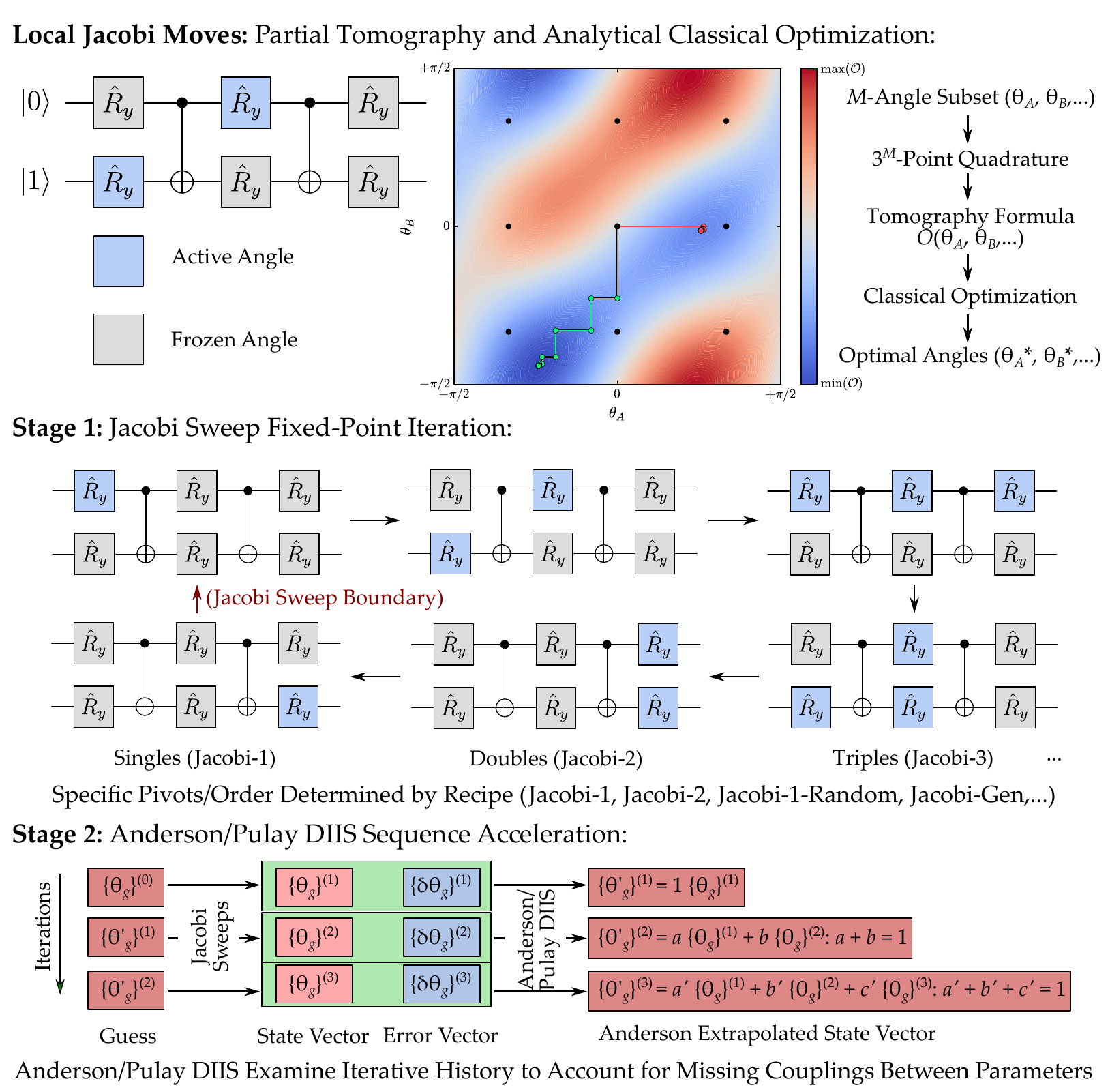}
\caption{
Overview of the family of Jacobi fixed point iteration plus Anderson/Pulay
DIIS sequence acceleration algorithms developed for variational quantum circuit
optimization in this work.
}
\label{fig:schem}
\end{center}
\end{figure*}

An overview of the family of Jacobi fixed point iteration plus Anderson/Pulay
DIIS sequence acceleration convergence algorithms developed herein is depicted
in Figure \ref{fig:schem}. In the top panel, we develop methodology to perform
tomography fitting of variational quantum circuits with respect to a subset of
variational parameters. This involves the development of a simple analytical
trigonometric form with problem-specific linear tomography parameters for the
partial tomography of the observable expectation value with respect to the
active cluster of parameters. The tomography parameters can be resolved by
sampling the observable expectation value at widely spaced Fourier quadrature
grid points, which are robust against cancellation error. Following the
tomography fitting, classical optimization can be used to determine the optimal
values of the active cluster of circuit parameters to arbitrary precsion. In the
middle panel, iterative sequences of these local moves are performed over
different clusters of active parameters, yielding a monotonically-converging
method that coarsely resembles classical Jacobi diagonalization.  In the bottom
panel, Anderson or Pulay DIIS sequence acceleration techniques are invoked to
examine the iterative history of the Jacobi sweeps and propose modified sets of
circuit parameters that promote cooperative optimization between the global set
of circuit parameters.

In Section II.A, we gradually build up the tomography formulae and Fourier
quadrature fitting procedures for various clusters and classes of quantum
circuits. Specifically, Section II.A.1 covers the simplest $M=1$ one-gate
tomography (which required 3 quadrature points to resolve), Section II.A.2
covers the $M=2$ double-gate tomography (which requires 9 quadrature points to
resolve), Section II.A.3 covers the general $M$-gate tomography (which requires
$3^M$ quadrature points to resolve), and Section II.A.4 covers the case
encountered in QAOA where there are $M$ clusters each of $G$ gates with the same
angle (which requires $(2G+1)^M$ quadrature points to resolve). In Section
II.B, we describe a number of possible recipes for selecting the sequence of
gate angle clusters to define each Jacobi sweep. In Section II.C, we describe
two variants of Anderson acceleration approaches to improve global convergence
properties of the approach: In Section II.C.1 we detail a traditional Anderson
recipe where the error vector is taken to be the difference in the state vector
between each move and the previous one, while in Section II.C.2 we detail a
method that more-closely resembles the Pulay DIIS procedure as commonly used in
SCF theory, where the gradient is used as the error vector. In Section II.D, we
comment on the similarity of Jacobi-1 and the Powell method.

In this work, we use the one-qubit $\hat R_y (\theta) \equiv e^{-i \theta \hat
Y}$ gate as a universal gate for parameter entry.

\subsection{VQE Tomography}

\subsubsection{One-Gate Tomography} 

Consider a general 1-$\hat R_y$-gate quantum circuit of the form encountered in
VQE,
\begin{widetext}
\begin{equation}
\Qcircuit @R=0.1em @C=0.3em @!R {
\lstick{|0_A\rangle}
 & \multigate{3}{\hat U}
 & \qw
 & \multigate{3}{\hat V}
 & \qw \\
\lstick{|0_B\rangle}
 & \ghost{\hat U}
 & \qw
 & \ghost{\hat V}
 & \qw \\
\lstick{|0_C\rangle}
 & \ghost{\hat U}
 & \gate{\hat R_y (\theta_A)}
 & \ghost{\hat V}
 & \qw \\
\lstick{|0_D\rangle}
 & \ghost{\hat U}
 & \qw
 & \ghost{\hat V}
 & \qw \\
}
\end{equation}
\end{widetext}
The corresponding observable expectation value is,
\begin{equation}
\mathcal{O} (\theta_A)
\equiv
\langle 0 | 
\hat U^\dagger
\hat R_{y}^{\dagger} (\theta_A)
\hat V^\dagger
\hat O
\hat V
\hat R_{y} (\theta_A)
\hat U
| 0 \rangle
\end{equation}
Here $\hat O$ is a Hermitian operator, and $\hat U$ and $\hat V$ are
unitary operators, all in the $2^N$-dimensional Hilbert space
for $N$ qubits. These matrices are problem-specific, and their construction is
determined classically before the invocation of the quantum algorithm. $\hat R_y
(\theta_A) \equiv e^{-i \theta_A \hat Y_A}$ is a one-qubit rotation gate acting
on an arbitrary qubit corresponding to index $A$.  The observable expectation
value $\mathcal{O} (\theta_A)$ is periodic for $\theta_A \in [-\pi / 2, + \pi /
2)$. By inspection, this has the definitional tomography,
\begin{equation}
\mathcal{O} (\theta_A)
\equiv
\alpha
+
\beta
\cos(2 \theta_A)
+
\gamma
\sin(2 \theta_A)
\end{equation}
Here, $\{ \alpha, \beta, \gamma \}$ are the ``tomography parameters,'' whose
values depend on the specifics of the problem (i.e., the operators $\hat U$,
$\hat V$, and $\hat O$). To determine the tomography parameters, we can define
the three-point Fourier quadrature,
\begin{equation}
\mathcal{O}^{-}
\equiv
\mathcal{O} (-\pi / 3)
, \
\mathcal{O}^{0}
\equiv
\mathcal{O} (0)
, \
\mathcal{O}^{+}
\equiv
\mathcal{O} (+\pi / 3)
\end{equation}
This has the matrix form,
\begin{equation}
\underbrace{
\left [
\begin{array}{ccc}
1 & -1/2 & -\sqrt{3} / 2 \\
1 & 1 &  0 \\
1 & -1/2 &  \sqrt{3} / 2 \\
\end{array}
\right ]
}_{\hat T_{(1)}}
\left [
\begin{array}{c}
\alpha \\
\beta \\
\gamma \\
\end{array}
\right ]
=
\left [
\begin{array}{c}
\mathcal{O}^{-} \\
\mathcal{O}^{0} \\
\mathcal{O}^{+} \\
\end{array}
\right ]
\end{equation}
The inverse is,
\begin{equation}
\hat T_{(1)}^{-1}
=
\frac{1}{3}
\left [
\begin{array}{ccc}
 1 & 1 &  1 \\
-1 & 2 & -1 \\
-\sqrt{3} & 0 &  \sqrt{3} \\
\end{array}
\right ]
\end{equation}
The gradient is,
\begin{equation}
\mathcal{G}_{A} (\theta_A)
\equiv
\pdiff{\mathcal{O} (\theta_A)}{\theta_A}
=
- 2 \beta \sin (2 \theta_A)
+ 2 \gamma \cos (2 \theta_A)
\end{equation}
The Hessian is,
\begin{equation}
\mathcal{H}_{AA} (\theta_A)
\equiv
\pdiff{^2 \mathcal{O} (\theta_A)}{\theta_A^2}
=
- 4 \beta \cos (2 \theta_A)
- 4 \gamma \sin (2 \theta_A)
\end{equation}
The stationary condition is,
\begin{equation}
\mathcal{G}_{A} (\theta_A)
=
0
\Rightarrow
\tan (2 \theta_A)
=
\gamma / \beta
\end{equation}
The minimal condition is,
\begin{equation}
\mathcal{H}_{AA} (\theta_A)
\geq
0
\end{equation}
And the analytical solution for the optimal angle is therefore,
\begin{equation}
\label{eq:theta-star-1}
\Rightarrow
\theta_A
=
\frac{1}{2}
\mathrm{arctan2}
\left (
\frac{-\gamma}{-\beta}
\right )
\end{equation}
This solution is unique - there is only one local minimum in 1D.

\subsubsection{Two-Gate Tomography} 

Now consider a general 2-$\hat R_y$-gate quantum circuit of the form encountered
in VQE,
\begin{widetext}
\begin{equation}
\Qcircuit @R=0.1em @C=0.3em @!R {
\lstick{|0_A\rangle}
 & \multigate{3}{\hat U}
 & \qw
 & \multigate{3}{\hat V}
 & \gate{\hat R_y (\theta_B)}
 & \multigate{3}{\hat W}
 & \qw \\
\lstick{|0_B\rangle}
 & \ghost{\hat U}
 & \qw
 & \ghost{\hat V}
 & \qw 
 & \ghost{\hat W}
 & \qw \\
\lstick{|0_C\rangle}
 & \ghost{\hat U}
 & \gate{\hat R_y (\theta_A)}
 & \ghost{\hat V}
 & \qw 
 & \ghost{\hat W}
 & \qw \\
\lstick{|0_D\rangle}
 & \ghost{\hat U}
 & \qw
 & \ghost{\hat V}
 & \qw
 & \ghost{\hat W}
 & \qw \\
}
\end{equation}
\end{widetext}
The corresponding observable expectation value is,
\[
\mathcal{O} (\theta_A, \theta_B)
\equiv
\langle 0 | 
\hat U^\dagger
\hat R_{y}^{\dagger} (\theta_A)
\hat V^\dagger
\hat R_{y}^{\dagger} (\theta_B)
\hat W^\dagger
\]
\begin{equation}
\times
\hat O
\hat W
\hat R_{y} (\theta_B)
\hat V
\hat R_{y} (\theta_A)
\hat U
| 0 \rangle
\end{equation}
By inspection, this has the tomography,
\[
\mathcal{O} (\theta_A, \theta_B)
\equiv
\alpha \alpha
+
\alpha \beta
\cos (2 \theta_B)
+
\alpha \gamma
\sin (2 \theta_B)
\]
\[
+
\beta \alpha
\cos (2 \theta_A)
+
\beta \beta
\cos (2 \theta_A)
\cos (2 \theta_B)
+
\beta \gamma
\cos (2 \theta_A)
\sin (2 \theta_B)
\]
\begin{equation}
\label{eq:tomography-2}
+
\gamma \alpha
\sin (2 \theta_A)
+
\gamma \beta
\sin (2 \theta_A)
\cos (2 \theta_B)
+
\gamma \gamma
\sin (2 \theta_A)
\sin (2 \theta_B)
\end{equation}
We can define a nine-point Fourier quadrature,
\[
\mathcal{O}^{--}
\equiv
\mathcal{O} (-\pi / 3, -\pi / 3)
, \
\mathcal{O}^{-0}
\equiv
\mathcal{O} (-\pi / 3, 0)
, \
\]
\[
\mathcal{O}^{-+}
\equiv
\mathcal{O} (-\pi / 3, +\pi / 3)
, \
\mathcal{O}^{0-}
\equiv
\mathcal{O} (0, -\pi / 3)
, \
\]
\[
\mathcal{O}^{00}
\equiv
\mathcal{O} (0, 0)
, \
\mathcal{O}^{0+}
\equiv
\mathcal{O} (0, +\pi / 3)
, \
\]
\[
\mathcal{O}^{+-}
\equiv
\mathcal{O} (+\pi / 3, -\pi / 3)
, \
\mathcal{O}^{+0}
\equiv
\mathcal{O} (+\pi / 3, 0)
, \
\]
\begin{equation}
\mathcal{O}^{++}
\equiv
\mathcal{O} (+\pi / 3, +\pi / 3)
, \
\end{equation}
This has the matrix form,
\begin{equation}
\hat T_{(2)}
\left [
\begin{array}{c}
\alpha \alpha \\ 
\alpha \beta \\ 
\alpha \gamma \\
\beta \alpha \\ 
\beta \beta \\ 
\beta \gamma \\
\gamma \alpha \\ 
\gamma \beta \\ 
\gamma \gamma \\
\end{array}
\right ]
=
\left [
\begin{array}{c}
\mathcal{O}^{--} \\
\mathcal{O}^{-0} \\
\mathcal{O}^{-+} \\
\mathcal{O}^{0-} \\
\mathcal{O}^{00} \\
\mathcal{O}^{0+} \\
\mathcal{O}^{+-} \\
\mathcal{O}^{+0} \\
\mathcal{O}^{++} \\
\end{array}
\right ]
\end{equation}
Here $\hat T_{(2)} \equiv \hat T_{(1)} \otimes \hat T_{(1)}$ and the
corresponding inverse is, $\hat T_{(2)}^{-1} = \hat T_{(1)}^{-1} \otimes \hat
T_{(1)}^{-1}$.

The gradient is,
\[
\mathcal{G}_{A} (\theta_A, \theta_B)
\equiv
\pdiff{\mathcal{O} (\theta_A, \theta_B)}{\theta_{A}}
\]
\[
=
- 2 \sin(2 \theta_A)
\left [
\beta \alpha
+
\beta \beta 
\cos (2 \theta_B)
+
\beta \gamma 
\sin (2 \theta_B)
\right ]
\]
\[
+
2 \cos(2 \theta_A)
\left [
\gamma \alpha
+
\gamma \beta 
\cos (2 \theta_B)
+
\gamma \gamma 
\sin (2 \theta_B)
\right ]
\]
\begin{equation}
\label{eq:grad-2}
\equiv
- 2 \sin(2 \theta_A)
\mathcal{O} (\theta_B | \beta_A)
+ 2 \cos(2 \theta_A)
\mathcal{O} (\theta_B | \gamma_A)
\end{equation}
And similar for $\mathcal{G}_{B} (\theta_A, \theta_B)$, by permutation of
indices.

The Hessian is,
\[
\mathcal{H}_{AA} (\theta_A, \theta_B)
\equiv
\pdiff{^2 \mathcal{O} (\theta_A, \theta_B)}
{\theta_{A}^2}
\]
\begin{equation}
=
- 4 \cos(2 \theta_A)
\mathcal{O} (\theta_B | \beta_A)
- 4 \sin(2 \theta_A)
\mathcal{O} (\theta_B | \gamma_A)
\end{equation}
And similar for $\mathcal{H}_{BB} (\theta_A, \theta_B)$, by permutation of
indices, and,
\[
\mathcal{H}_{AB} (\theta_A, \theta_B)
\equiv
\pdiff{^2 \mathcal{O} (\theta_A, \theta_B)}
{\theta_{A} \partial \theta_{B}}
\]
\begin{equation}
=
- 2 \sin(2 \theta_A)
\pdiff{\mathcal{O} (\theta_B | \beta_A)}{\theta_B}
+ 2 \cos(2 \theta_A)
\pdiff{\mathcal{O} (\theta_B | \gamma_A)}{\theta_B}
\end{equation}

The stationary condition is,
\[
\mathcal{G}_{A} (\theta_A, \theta_B)
= 0
\]
\begin{equation}
\label{eq:stationary-2A}
\Rightarrow
\tan (2 \theta_A)
=
\frac{
\gamma \alpha
+
\gamma \beta 
\cos (2 \theta_B)
+
\gamma \gamma 
\sin (2 \theta_B)
}{
\beta \alpha
+
\beta \beta 
\cos (2 \theta_B)
+
\beta \gamma 
\sin (2 \theta_B)
}
\end{equation}
and,
\[
\mathcal{G}_{B} (\theta_A, \theta_B)
= 0
\]
\begin{equation}
\label{eq:stationary-2B}
\Rightarrow
\tan (2 \theta_B)
=
\frac{
\alpha \gamma
+
\beta \gamma
\cos (2 \theta_A)
+
\gamma \gamma
\sin (2 \theta_A)
}{
\alpha \beta 
+
\beta \beta
\cos (2 \theta_A)
+
\gamma \beta
\sin (2 \theta_A)
}
\end{equation}
The minimal condition is,
\begin{equation}
\label{eq:minimal}
\mathcal{H}
\geq 0
\end{equation}
Note that this last expression means that the eigenspectrum of the Hessian
should be positive at the stationary point to guarantee a local minimum. Also
note that multiple local minima are sometimes present in the observables for
$\geq 2$-gate tomography (we have empirically seen $1\times$ or $2\times$ local
minima in 2-gate tomography).

We have not, as yet, determined an analytical solution for the optimal angles
$\theta_A$ and $\theta_B$. However, after the tomography fitting has been
performed, it is straightforward to classically optimize the angles to close to
the machine precision on the analytical tomography surface. One particularly
simple and robust procedure we have developed is to solve for the analytical
optimal value of $\theta_{A}$ while $\theta_{B}$ is frozen (via applying the
recipe of Equation \ref{eq:theta-star-1} to Equation \ref{eq:stationary-2A}) and
then to repeat to analytically optimize $\theta_{B}$ while $\theta_{A}$ is
frozen. This procedure can be applied iteratively, and is guaranteed to converge
monotonically toward a local minimum.  We refer to this approach as the
classical Jacobi-1 optimization approach (``1'' for one single gate angle at a
time). The monotonic convergence property makes the Jacobi optimization
procedure particularly robust relative to gradient/Hessian-based methods such as
Newton-Raphson or L-BFGS, which often fail or converge slowly for this objective
function due to the indefinite or negative nature of the Hessian in large
patches of the parameter space - a phenomenon for variational quantum observable
expectation values that has previously been referred to as the ``barren
plateaus'' issue.\cite{McClean:2018:X}  Note that there are two caveats with
the Jacobi procedure: (1) the procedure may converge to spurious local minima
and (2) the procedure converges rather slowly if there is high covariance of
the observable expectation value $\mathcal{O} (\theta_A, \theta_B)$ with
respect to the two parameters $\theta_A$ and $\theta_B$. To combat (1), the
classical tomography formula can be sampled with a medium-density rectilinear
grid (e.g., a spacing of $\Delta \theta = \pi / 8$) and the Jacobi-1 iteration
seeded from the global minimum on the medium-density grid. To combat (2), one
can switch to classical L-BFGS or full classical Newton-Raphson once the Jacobi
optimization procedure has reached the neighborhood of a quadratic minimum.

\begin{figure}[h!]
\begin{center}
\includegraphics[width=3.4in]{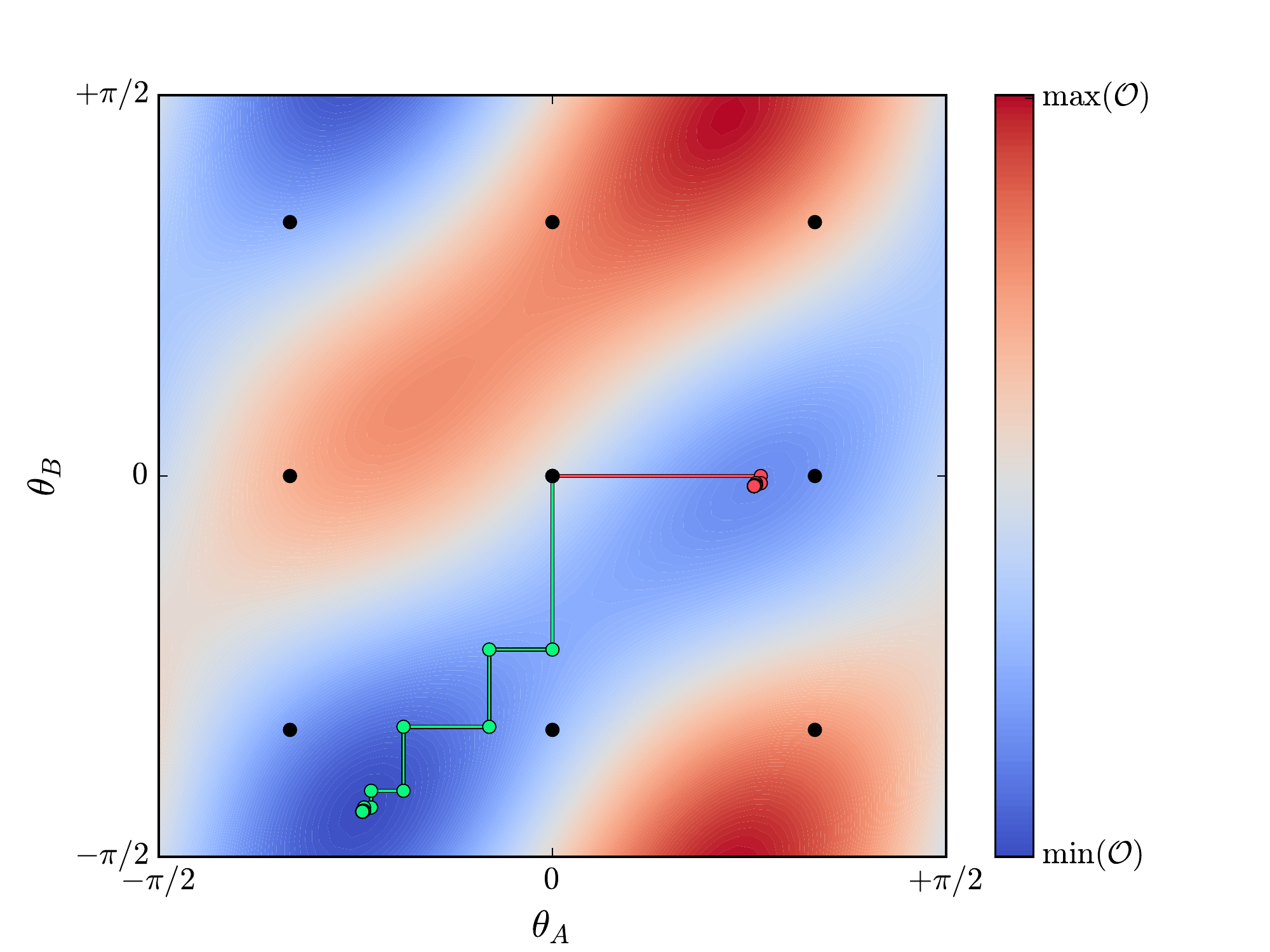}
\caption{Schematic of 9-point quadrature and classical Jacobi-1 optimization
procedure for two-gate tomography. The observable expectation value $\mathcal{O}
(\theta_A, \theta_B)$ is an example computed on a dense reference grid and
depicted as the filled contour map. Sampling the observable on the 9-point
quadrature grid depicted as filled black circles is sufficient to recover the
analytical tomography formula of Equation \ref{eq:tomography-2}.
Subsequently, the optimal angles $\theta_A$ and $\theta_B$ can be found by
classical optimization techniques, e.g., by the classical Jacobi-1 approach
depicted in the green and pink lines. However, care must be taken to avoid
spurious local minima which might be present, such as the solution found by the
pink classical Jacobi-1 optimization.}
\label{fig:schem-2d}
\end{center}
\end{figure}

A schematic of the 9-point quadrature and classical Jacobi-1
optimization procedure for two-gate tomography is depicted in Figure
\ref{fig:schem-2d}.

\subsubsection{$M$-Gate Tomography} 

Now consider a general $M$-$\hat R_y$-gate quantum
circuit of the form encountered in VQE,
\begin{widetext}
\begin{equation}
\begin{array}{l}
\Qcircuit @R=0.1em @C=0.3em @!R {
\lstick{|0_A\rangle}
 & \multigate{3}{\hat U}
 & \qw
 & \multigate{3}{\hat V}
 & \gate{\hat R_y (\theta_B)}
 & \multigate{3}{\hat W}
 & \qw \\
\lstick{|0_B\rangle}
 & \ghost{\hat U}
 & \qw
 & \ghost{\hat V}
 & \qw 
 & \ghost{\hat W}
 & \qw \\
\lstick{|0_C\rangle}
 & \ghost{\hat U}
 & \gate{\hat R_y (\theta_A)}
 & \ghost{\hat V}
 & \qw 
 & \ghost{\hat W}
 & \qw \\
\lstick{|0_D\rangle}
 & \ghost{\hat U}
 & \qw
 & \ghost{\hat V}
 & \qw
 & \ghost{\hat W}
 & \qw \\
}
\end{array}
\ldots
\begin{array}{l}
\Qcircuit @R=0.1em @C=0.3em @!R {
 & \qw 
 & \multigate{3}{\hat Z}
 & \qw \\
 & \qw 
 & \ghost{\hat Z}
 & \qw \\
 & \qw 
 & \ghost{\hat Z}
 & \qw \\
 & \gate{\hat R_y (\theta_Z)}
 & \ghost{\hat Z}
 & \qw \\
}
\end{array}
\end{equation}
\end{widetext}
The corresponding observable expectation value is,
\[
\mathcal{O} (\theta_A, \theta_B, \ldots, \theta_Z)
\equiv
\]
\[
\langle 0 | 
\hat U^\dagger
\hat R_{y}^{\dagger} (\theta_A)
\hat V^\dagger
\hat R_{y}^{\dagger} (\theta_B)
\hat W^\dagger
\ldots
\hat R_{y}^{\dagger} (\theta_Z)
\hat Z^\dagger
\]
\begin{equation}
\times
\hat O
\hat Z
\hat R_{y} (\theta_Z)
\ldots
\hat W
\hat R_{y} (\theta_B)
\hat V
\hat R_{y} (\theta_A)
\hat U
| 0 \rangle
\end{equation}
Note that the alphabetical labeling $\theta_A, \theta_B, \ldots, \theta_Z$ is
merely illustrative - there can be more or less than 26 gates depending on $M$.
This has the $3^M$-parameter tomography,
\[
\mathcal{O} (\theta_A, \theta_B, \ldots, \theta_Z)
=
\sum_{\vec I}
c_{\vec I}
\prod_{i_D \in I}
\phi_{i_D} (\theta_D)
\]
Where $\vec I$ ranges over the set of trinary strings of length $M$, e.g.,
$000$, $001$, $002$, $010$, $011$, $012$, $020$, $021$, $022$, $100$, etc, and
the basis functions are $\phi_{0} (\theta) \equiv 1$, $\phi_{1} (\theta) \equiv
\cos(2 \theta)$, and $\phi_{2} (\theta) \equiv \sin(2 \theta)$. 

The tomography coefficients $\{ c_{\vec I} \}$ can be computed from a
$3^M$-point quadrature grid consisting of a Cartesian grid of $\{ - \pi / 3, 0,
+ \pi / 3 \}$ in each angle. The transfer matrix is $\hat T_{(M)} \equiv
\bigotimes_{M} \hat T_{(1)}$, and the corresponding inverse is $\hat
T_{(M)}^{-1} \equiv \bigotimes_{M} \hat T_{(1)}^{-1}$.

The gradient, Hessian, stationary conditions, and minimal conditions all follow
straightforwardly from multi-dimensional extensions of Equations \ref{eq:grad-2}
to \ref{eq:minimal}, e.g.,
\[
\mathcal{G}_{A} (\theta_A, \theta_B, \ldots, \theta_Z)
\equiv
\pdiff{\mathcal{O} (\theta_A, \theta_B, \ldots, \theta_Z)}{\theta_{A}}
\]
\[
=
- 2 \sin(2 \theta_A)
\mathcal{O} (\theta_B, \ldots, \theta_Z | \beta_A)
\]
\begin{equation}
+ 2 \cos(2 \theta_A)
\mathcal{O} (\theta_B, \ldots, \theta_Z | \gamma_A)
\end{equation}

\subsubsection{QAOA Tomography}

In QAOA,\cite{Farhi:2014:Quantum} we often encounter a
generalization/simplification where the angles of multiple $\hat R_y$ gates are
pinned together. E.g., for an $M$-stage quantum circuit, where each stage $D$
has $G_D$ $\hat R_y (\theta_D)$ gates,
\begin{widetext}
\begin{equation}
\underbrace{
\begin{array}{l}
\Qcircuit @R=0.1em @C=0.3em @!R {
\lstick{|0_A\rangle}
 & \multigate{3}{\hat U}
 & \qw
 & \multigate{3}{\hat V}
 & \gate{\hat R_y (\theta_A)}
 & \multigate{3}{\hat W}
 & \qw \\
\lstick{|0_B\rangle}
 & \ghost{\hat U}
 & \qw
 & \ghost{\hat V}
 & \qw 
 & \ghost{\hat W}
 & \qw \\
\lstick{|0_C\rangle}
 & \ghost{\hat U}
 & \gate{\hat R_y (\theta_A)}
 & \ghost{\hat V}
 & \qw 
 & \ghost{\hat W}
 & \qw \\
\lstick{|0_D\rangle}
 & \ghost{\hat U}
 & \qw
 & \ghost{\hat V}
 & \qw
 & \ghost{\hat W}
 & \qw \\
}
\end{array}
}_{
G_{A} \ \mathrm{\hat R_y (\theta_A)} \ \mathrm{Gates}
}
\ldots
\underbrace{
\begin{array}{l}
\Qcircuit @R=0.1em @C=0.3em @!R {
 & \qw 
 & \multigate{3}{\hat Y}
 & \qw 
 & \multigate{3}{\hat Z}
 & \qw \\
 & \gate{\hat R_y (\theta_Z)}
 & \ghost{\hat Y}
 & \qw 
 & \ghost{\hat Z}
 & \qw \\
 & \qw 
 & \ghost{\hat Y}
 & \qw 
 & \ghost{\hat Z}
 & \qw \\
 & \qw 
 & \ghost{\hat Y}
 & \gate{\hat R_y (\theta_Z)}
 & \ghost{\hat Z}
 & \qw \\
}
\end{array}
}_{
G_{Z} \ \mathrm{\hat R_y (\theta_Z)} \ \mathrm{Gates}
}
\end{equation}
\end{widetext}
The corresponding observable expectation value is,
\[
\mathcal{O} (\theta_A, \ldots, \theta_Z)
\equiv
\]
\[
\langle 0 | 
\hat U^\dagger
\hat R_{y}^{\dagger} (\theta_A)
\hat V^\dagger
\hat R_{y}^{\dagger} (\theta_A)
\hat W^\dagger
\ldots
\hat R_{y}^{\dagger} (\theta_Z)
\hat Y^\dagger
\hat R_{y}^{\dagger} (\theta_Z)
\hat Z^\dagger
\]
\begin{equation}
\times
\hat O
\hat Z
\hat R_{y} (\theta_Z)
\hat Y
\hat R_{y} (\theta_Z)
\ldots
\hat W
\hat R_{y} (\theta_A)
\hat V
\hat R_{y} (\theta_A)
\hat U
| 0 \rangle
\end{equation}
This has the $\prod_{D} (2 G_D + 1)$-parameter tomography [$(2 G + 1)^M$ if
$G_D$ is independent of $D$],
\begin{equation}
\mathcal{O} (\theta_A, \ldots, \theta_Z)
=
\sum_{\vec I}
c_{\vec I}
\prod_{i_D \in \vec I}
\phi_{i_D} (\theta_D)
\end{equation}
Where each digit of $\vec I$ ranges from $-G_D$ to $+G_D$ (inclusive). The basis
functions are are $\sin(G_D \cdot 2 \theta)$, \ldots, $\sin(2 \cdot 2 \theta)$,
$\sin(1 \cdot 2 \theta)$, $1$, $\cos (1 \cdot 2 \theta)$, $\cos(2 \cdot 2
\theta)$, \ldots, $\cos(G_D \cdot 2 \theta)$.

Two technical notes with the circuit above: (1) in QAOA, often the interstitial
operators $\hat U$, $\hat V$, \ldots are the identity, e.g., within commuting
layers of 1-qubit driver terms and (2) above we have drawn the $\theta_A$ and
$\theta_Z$ stages as disjoint, but they may be interleaved without changing the
analysis.

For this case, a Cartesian product of $(2 G_D + 1)$-point Fourier grids is
sufficient to resolve the tomography. The task is simply to redefine the grid,
transfer matrix, and transfer matrix inverse.  For general $G_D$, the Fourier
grid is,
\begin{equation}
\theta_{p}
=
\frac{
(p + 1) \pi
}{
2 G_D + 1
}
-
\frac{\pi}{2}
, \ 
p \in [0, 2 G_D + 1)
\end{equation}
The transfer matrix is,
\begin{equation}
T_{ip}^{G}
\equiv
\phi_{i} (\theta_p)
=
\left [
\begin{array}{c}
\sin (2 i \theta_p) \\ \hline
1 \\ \hline
\cos (2 i \theta_p)  \\
\end{array}
\right ]
\end{equation}
As before, the $M$-stage transfer matrix is $\hat T_{(M)} \equiv
\bigotimes_{D}^{M} \hat T_{(1)}^{G_D}$, and the corresponding inverse is $\hat
T_{(M)}^{-1} \equiv \bigotimes_{D}^{M} \hat T_{(1)}^{G_D,-1}$.

The gradient and Hessian are easily computed along the lines of the approach
used in the previous section, with specific partial derivatives of the basis
functions $\phi_{i} (\theta_p)$.

\begin{figure}[h!]
\begin{center}
\includegraphics[width=3.4in]{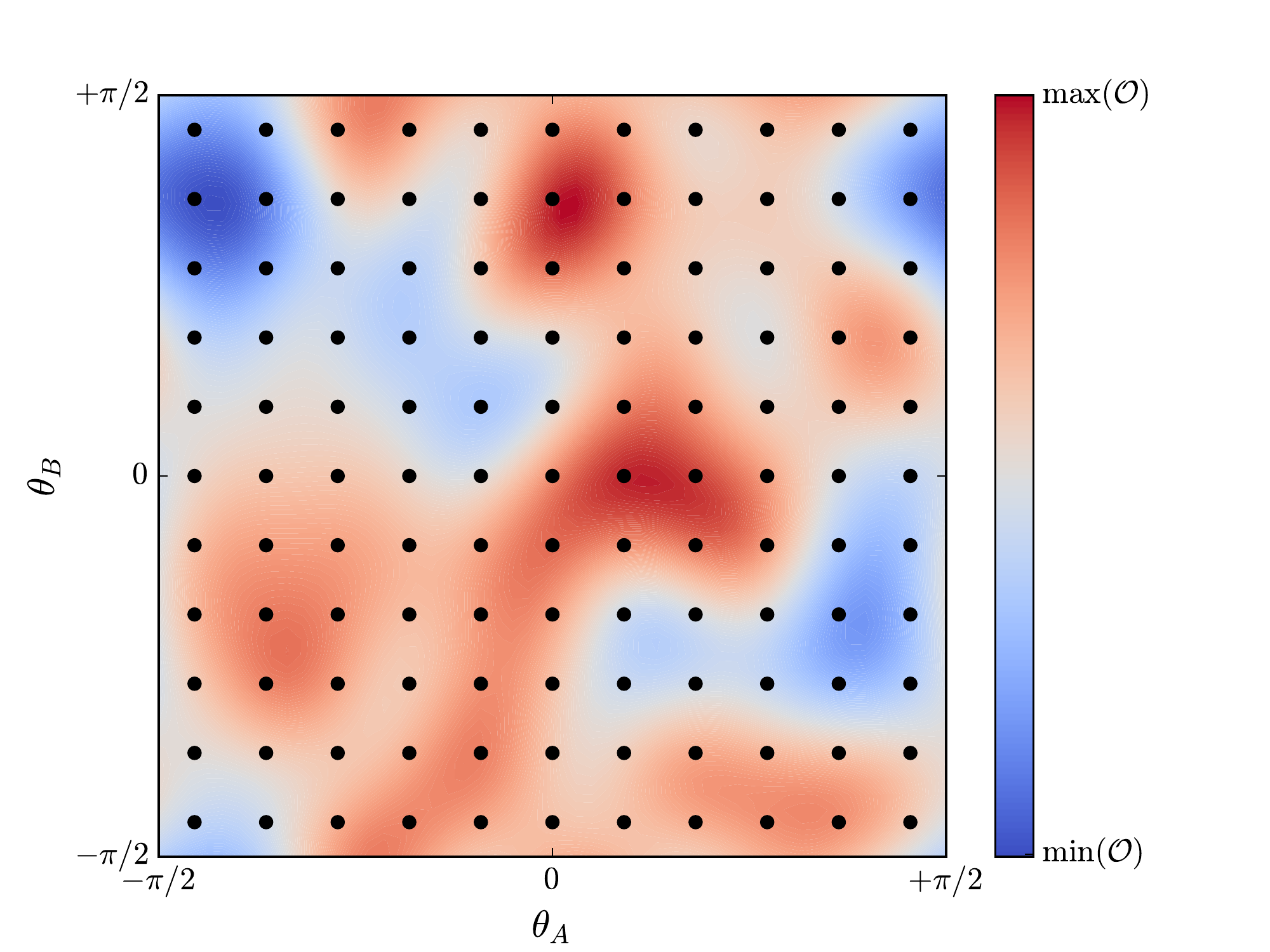}
\caption{Example of using Fourier quadrature for QAOA tomography ($M=2$,
$G_D=5$).  $(2\cdot5+1)^2 = 121$ quadrature points are required to analytically
resolve the tomography coefficients.}
\label{fig:test-qaoa}

\end{center}
\end{figure}

An example Fourier quadrature resolution of a QAOA-type tomography function is
depicted in Figure \ref{fig:test-qaoa}.

\subsubsection{Tomography Formula Verification}

The tomography formulae and quadrature recipes developed above were verified by
dense Fourier grid comparison to randomly-generated quantum circuits (i.e.,
random $\hat O$ and $\hat U$, $\hat V$, \ldots) for up to $M=5$ and up to to
$G=5$. All test cases exhibit a relative deviation between the analytical
circuit simulation result and the quadrature-fitted tomography formula of
$\sim100 \epsilon$, where $\epsilon \equiv 2.2\times10^{-16}$ is the
double-precision machine epsilon, verifying the that quadrature-based tomography
fitting is analytical to close to the machine precision.

In closing this section, we make two observations regarding the Fourier
quadrature resolution of the tomography parameters. The first observation is
that the quadrature points are widely spaced in $\theta$, and the tomography
coefficients are resolved by the transfer matrix inverse with coefficients that
are nearly unity. This means that roughly the same statistical convergence of
the observable expectation value at each quadrature grid point is required to obtained a
given absolute accuracy in the observable across the full tomography formula as
is required to obtain the same absolute accuracy in the observable at a specific
point. This is in marked contrast to, e.g., finite difference derivatives, where
the observable must be resolved to much higher precision at the stencil
grid points to obtain a given accuracy in the approximated derivative,
due to subtractive cancelation. Since we are free to analytically differentiate
the tomography formula, we obtain a recipe for the analytical gradient of the
observable expectation which has the same number of required observable
expectation value evaluations as the second-order symmetric finite difference
formula, but with markedly reduced precision requirements. Tacitly, for a
single-gate example,
\begin{equation}
\mathcal{G}_A (\theta_A)
\equiv
\pdiff{\mathcal{O}}{\theta_A}
=
\frac{2}{\sqrt{3}}
\mathcal{O} (\theta_A + \pi / 3)
-
\frac{2}{\sqrt{3}}
\mathcal{O} (\theta_A - \pi / 3)
\end{equation}
\begin{equation}
=
\lim_{h \rightarrow 0}
\left [
\frac{1}{2h}
\mathcal{O} (\theta_A + h)
-
\frac{1}{2h}
\mathcal{O} (\theta_A - h)
\right ]
\end{equation}
Here the top formula relies on the tomography formula resolved by Fourier
quadrature, while the bottom formula uses the second-order symmetric finite
difference formula. Both require $2 P$ observable evaluations to compute the
complete gradient for $P$ gates, but the upper formula has markedly lower
precision requirements at each grid point. Similar formulae occur for the
Hessian and higher-order derivatives - in each case the number of required
evaluations is the same as in second-order finite difference, but the wide
spacing of the quadrature grid points reduce the precision requirements
at each grid point. The only potential downside is that the computation
of directional derivatives (i.e., the gradient projected along a given linear
combination of circuit angles) requires the full $2 P$ observable evaluations
with the Fourier quadrature recipe, but generally requires $\mathcal{O}(1)$
observable evaluations with finite difference. The second observation is that the Fourier
quadrature selected for the tomography fitting in this study is not the only
valid choice of quadrature grid: any three distinct quadrature points would
produce a non-singular transfer matrix that could analytically determine the
tomography coefficients. For example, we have previously used the modified
quadrature $\{ -\pi / 4, 0, + \pi / 4 \}$, and have also been able to resolve
the analytical tomography formulae to $\sim 100 \epsilon$, as expected. A final
note is that a post-3-point quadrature (e.g., an $E$-point quadrature or a
stochastic sampling along $\theta_A$) might prove to be useful for error
mitigation during tomography fitting on noisy quantum hardware. Here, additional
information from the extended grid recipe could identify statistically
significant differences from the expected tomography formula, which is
definitionally noise and can be excluded by the extended tomography fitting.

\subsection{Jacobi-Type Local Fixed-Point Iteration}

At this point, the quadrature grid points, transformation to tomography
coefficients, and characteristics such as gradients and Hessians can all be
determined. What remains is to use these iterated partial tomography
measurements to classically optimize VQE or QAOA quantum circuit parameters, as
sketched in the middle panel of Figure \ref{fig:schem}.  There seem to be many
possible recipes involving maximum cluster size $M$, mixings of clusters of
different sizes, cluster order, cluster order randomization, and consideration
of circuit layout in selecting clusters and order. Here, we describe a few
fairly obvious recipes,
\begin{description}
\item [Jacobi-1:] All $M=1$ single angles, in a definite order during each
Jacobi sweep. This method requires a 3-point Fourier quadrature for the
tomography at each angle, leading to $3P$ total observable evaluations for $P$
gates (reducible $2P+1$ if the central point of each quadrature is inferred from
the tomography of the previous angle). This is complementary with the $2P$
observable evaluations required to compute the total circuit gradient in
gradient-based methods, though note that the observations in gradient-based
methods are parallelizable/pipelineable, while the observations in Jacobi-based
measurements are usually serialized.
\item [Jacobi-2:] All $M=2$ pairs of angle, in a definite order during each
Jacobi sweep. This method requires a 9-point Fourier quadrature for the
tomography at each angle pair, leading to $9 P (P + 1) / 2$ total observable
evaluations for $P$ gates (prefactor reducible if the values at certain points
are inferred from tomography fittings for other angles, in certain orders of
angle pairs).  This is complementary with the $4 P (P+1) / 2 + 2 P + 1$
observable evaluations required to compute the total circuit Hessian for full
Newton-Raphson-based optimization methods.
\item [Jacobi-Gen:] A generalized approach, with a user-specified sequence of
clusters of angles is optimized in a given order. This requires $3^M$ observable
evaluations for each $M$-gate cluster. The selection of such clusters is
presently an art, but could be guided by considerations of spatial or temporal
geometry within the quantum circuit, explicit computations to determine angle
clusters with strong coupling, or other methods. The heuristic choice(s)
selected for the numerical tests in the present manuscript are detailed below.
\item [Jacobi-1-Rand:] Jacobi-1, but with the order of angles shuffled randomly.
\item [Jacobi-2-Rand:] Jacobi-2, but with the order of angles shuffled randomly.
\item [Jacobi-Gen-Rand:] Jacobi-Gen, but with the order of angles shuffled randomly.
\end{description}
Many other possible Jacobi sweep recipes surely must exist - their determination
is a worthy topic of future study.

\subsection{Anderson-Type Global Convergence Acceleration}

In practice, we have found that one additional step can sometimes substantially
accelerate the global convergence of the Jacobi-type algorithm developed above,
at no additional cost (i.e., no additional quantum measurements, and very few
classical operations): Anderson-type sequence acceleration. The Anderson method
is arguably an unusual numerical methods technique that is often described as
more of a recipe than an algorithm. It was developed by
Anderson\cite{Anderson:1965:547} in 1965 as a pragmatic means of accelerating
slowly-converging sequences, and was independently discovered as the direct
inversion of the iterative subspace (DIIS) by Pulay\cite{Pulay:1980:393} in
1980 (and furthered\cite{Pulay:1982:556} in 1982) in the domain of electronic
structure theory. Anderson acceleration resembles Krylov linear algebraic
methods such as Lanczos, Arnoldi, GMRES, or Davidson-Liu, or the low-rank
Hessian update quasi-Newton optimization methods such as Broyden, DFP, or
L-BFGS in that it examines the iterative sequence and provides adjustive
predictions to the sequence to accelerate convergence. However, the formal
details of how Anderson acceleration works are mathematically rather murky -
there are few guiding principles on how and why the method works, and this is
an area of ongoing study. Part of the reason for this is that Anderson is
actually a class of methods with many possible variations and applications.
There have been several formal efforts that have shown that certain variations
and applications of Anderson reduce to GMRES (for application to linear
solutions),\cite{walker2011anderson} to Arnoldi iteration (for application to
eigenstate solutions),\cite{walker2011anderson} or to a multi-secant-type
method (for application to nonlinear root finding),\cite{fang2009two} and have
probed the formal convergence properties in those areas. 

The basic idea of Anderson is, for a given set of nonlinear equations $f(\vec
\theta) = 0$ (the ``vector'' symbol indicates the dimension of the parameter
space), one is given an iterative sequence over iteration index $i$ of state
vectors $\vec \theta^{i}$ and error vectors $\vec \epsilon^{i} \equiv \vec
\theta^i - \vec \theta$ so that $\vec \theta^{i} = \vec \theta + \vec
\epsilon^i$. Anderson acceleration posits that at iteration $k$, an improved
state vector $\vec \theta^{\prime k}$ can replace $\vec \theta^{k}$, and is
composed of a linear combination of the current history of iterative state
vectors,
\begin{equation}
\vec \theta^{\prime k}
\equiv
\sum_{i}
c_{i}^k
\vec \theta^{i}
\end{equation}
The corresponding improved error vector is,
\begin{equation}
\vec \epsilon^{\prime k}
\equiv
\sum_{i}
c_{i}^k
\vec \epsilon^{i}
\end{equation}
Anderson acceleration proposes that the coefficients $c_i^k$ be chosen to
minimize the square of the 2-norm of $\vec \epsilon^{\prime k}$,
\begin{equation}
O^k (c_i)
=
\sum_{ij}
c_i^k
c_j^k
\vec \epsilon^{i}
\cdot
\vec \epsilon^{j}
\end{equation}
subject to the normalization condition,
\begin{equation}
\sum_{i}
c_i^k
= 
1
\end{equation}
In the limit of a sufficiently large iterative space to span the vector space
for $\vec \theta$, this is guaranteed to produce the desired result of
$\theta^{\prime k} = \theta^{k}$, and will provide a least-squares error
approximation in a less-complete limit. An analytical solution for the
linearly-constrained least-squares problem is easily obtained,
\begin{equation}
\left [
\begin{array}{ccc|c}
B_{11} & \ldots & B_{1k} & -1 \\
\vdots & \ddots & \vdots & \vdots \\
B_{k1} & \ldots & B_{kk} & -1 \\
\hline
-1 & \ldots & -1 & 0 \\
\end{array}
\right ]
\left [
\begin{array}{c}
c_{1}^k \\
\vdots \\
c_{k}^k \\
\hline
\lambda \\
\end{array}
\right ]
=
\left [
\begin{array}{c}
0 \\
\vdots \\
0 \\
\hline
-1 \\
\end{array}
\right ]
\end{equation}
Here $B_{ij} \equiv \vec \epsilon^i \cdot \vec \epsilon^j$ and $\lambda$ is a
Langrange multiplier corresponding to the normalization constraint. The
numerical cost of predicting $\theta^{\prime k}$ from the iterative history is
minimal (and all classical): only the inner-products in $B_{ij}$, a matrix
inversion of the dimension of $k+1$, and the vector addition to form
$\theta^{\prime k}$ are required.

The cognizant reader will have noticed that the above manipulations would all be
for naught if we actually had the iterative history of the error vector $\vec
\epsilon^k$: at any iteration (including the first iteration!), we could simply
obtain the full solution as $\vec \theta = \vec \theta_k - \vec \epsilon_k$. In
practice, Anderson acceleration replaces the iterative history of the exact
error vector $\vec \epsilon^{k}$ with
a proxy quantity $\tilde \vec \epsilon^{k}$ that is pragmatically chosen to
approximate a scaled error vector. For instance, the displacement of the
parameters during a fixed-point iterative move $\delta \vec \theta^{k} \equiv
\vec \theta^{k} - \vec \theta^{\prime k-1}$ (the original Anderson\cite{Anderson:1965:547} and Pulay
DIIS\cite{Pulay:1980:393} proposals) may be effective slowly converging fixed-point series which are
making large numbers of small moves in roughly the same direction (e.g., the
convergence of the classical Jacobi-1 procedure for the $M=2$ example in Figure
\ref{fig:schem-2d}).  Alternatively, an approximately preconditioned gradient
may prove more useful for accelerating nonlinear optimization procedures posed
as gradient root finding or linear solve procedures posed as residual zero
finding - this procedure was used in Pulay's second DIIS paper in 1982,\cite{Pulay:1982:556} and is
industry standard for converging classical SCF equations. 

Note that the invocation of Anderson or Pulay DIIS sequence acceleration
techniques invalidate the monotonic convergence property of the Jacobi
diagonalization-type procedure developed above. However, we note that
Anderson/Pulay DIIS are robust methods - if a bad ``move'' is proposed, it will
usually have a large associated error vector, and therefore its DIIS coefficient
will be small. 

\subsubsection{Anderson-Style Acceleration}

There are some implementation subtleties regarding the order of Jacobi and
Anderson steps, the initiation of the iterative sequence, and the timing and
history of the Anderson subspace. For Anderson-style acceleration, the explicit
sequence is,
\begin{equation}
\vec \theta^{0}
\stackrel{\mathrm{Jacobi}}{\rightarrow}
\vec \theta^{1}
\stackrel{\mathrm{DIIS}[\vec \theta^{1}, \delta \vec \theta^{1} \equiv \vec
\theta^1 - \vec \theta^0]}{\rightarrow}
\end{equation}
\[
\vec \theta^{\prime 1}
\stackrel{\mathrm{Jacobi}}{\rightarrow}
\vec \theta^{2}
\stackrel{\mathrm{DIIS}[\vec \theta^{2}, \delta \vec \theta^{2} \equiv \vec
\theta^2 - \vec \theta^{\prime 1}]}{\rightarrow}
\]
\[
\vec \theta^{\prime 2}
\stackrel{\mathrm{Jacobi}}{\rightarrow}
\vec \theta^{3}
\stackrel{\mathrm{DIIS}[\vec \theta^{3}, \delta \vec \theta^{3} \equiv \vec
\theta^3 - \vec \theta^{\prime 2}]}{\rightarrow}
\]
\[
\vdots
\]
Here $\mathrm{DIIS}[\vec \theta, \vec \epsilon]$ is a subroutine call that adds
state vector $\vec \theta$ and error vector $\vec \epsilon$ to the iterative
history, and then returns an extrapolated result for $\vec \theta$ from the
current contends of the iterative history. 

\subsubsection{Pulay-Style DIIS Acceleration}

Similarly for Pulay-style DIIS as commonly implemented in SCF,
\begin{equation}
\vec \theta^{0}
\stackrel{\mathrm{DIIS}[\vec \theta^{0}, \vec G (\theta^0)]}{\rightarrow}
\vec \theta^{\prime 0}
\stackrel{\mathrm{Jacobi}}{\rightarrow}
\end{equation}
\[
\vec \theta^{1}
\stackrel{\mathrm{DIIS}[\vec \theta^{1}, \vec G (\theta^1)]}{\rightarrow}
\vec \theta^{\prime 1}
\stackrel{\mathrm{Jacobi}}{\rightarrow}
\]
\[
\vec \theta^{2}
\stackrel{\mathrm{DIIS}[\vec \theta^{2}, \vec G (\theta^2)]}{\rightarrow}
\vec \theta^{\prime 2}
\stackrel{\mathrm{Jacobi}}{\rightarrow}
\]
\[
\vdots
\]
Here $\vec G (\vec \theta)$ is the gradient vector
$\pdiff{\mathcal{O}}{\theta_g}$ at the current parameter set.  With some minimal
logical statements, these can be written in one monolithic Jacobi+Anderson code.

\subsection{Relationship with Powell's Method}

Consideration of the methods developed above with existing classical
optimization algorithms reveals an interesting comparison of the Jacobi-1 method
with a closely related algorithm: Powell's gradient-free optimization
method.\cite{Powell:1964:155} Each iteration of Powell's method involves
sweeping over a set of search directions (a set of generally non-orthogonal
normal vectors in the parameter space, with the set dimension being equal to the
number of parameters) and sequentially performing a bidirectional linesearch
along each search direction to find a local minimum. At the end of each
iteration, the total displacement of the iteration is added as a new search
direction, and the previous search direction that contributed the most to the
total displacement over the iteration is discarded. The first iteration of
Powell and Jacobi-1 are identical if the standard normal vectors in the basis of
circuit parameters are used as the initial search directions for Powell.
Subsequently, Powell is able to avoid the convergence stagnation that will be
observed shortly for DIIS-free Jacobi-1 by searching along directions that are
linear combinations of multiple circuit parameters (whereas Jacobi-1 is always
constrained to search along only individual circuit parameters). The penalty for
the flexibility of Powell is that a bidirectional linesearch must be performed
along each search direction, generally requiring a modest number ($\sim 12-14$)
of observable expectation values for each search direction. 

Jacobi-1 augmented with Anderson or Pulay DIIS might offer a method that is
correspondent and competitive with Powell: here, the number of observables per
Jacobi-1 iteration is strictly $3M$ ($2M + 1$ with tomography formula reuse),
and the raw Jacobi-1 constraint to rectilinear search directions along circuit
single circuit parameters is mitigated by the DIIS extrapolation, allowing one
to move ``diagonally'' through the parameter space. Below we will see that the
early convergence history of Jacobi-1-Anderson and Jacobi-1-DIIS are remarkably
similar to Powell, though the latter method requires several times more
observables per iteration to carry out linesearches.

\section{Computational Details}

The Fourier quadrature grid based tomography fitting procedure and
Jacobi+Anderson optimization algorithms were implemented for MC-VQE in our
in-house quantum simulator package \textsc{Quasar}. For the purposes of this
study, noise channels and statistical errors are not modeled: the observable
expectation values are computed by double-precision contraction of the simulated
qubit wavefunctions to the relevant Pauli density matrices. This directly probes
the characteristics  of the algorithms in the absence of noise, allowing for analysis
of both the initial convergence behavior (in the statistical and NISQ-limited
hardware regime) and the terminal/global convergence behavior (in the
tightly-converged limit, relevant for classical benchmarking). Future studies
will investigate the robustness of these algorithms under statistical and device
noise channels, on simulated and physical hardware.

To demonstrate the characteristics of the involved algorithms, we have selected
an ``easy'' and ``hard'' VQE optimization test case from our recently-developed
MC-VQE+AIEM methodology. In both cases, we consider an $N=4$ monomer/qubit
system of a linearly arranged set of chromophores from the B850 ring of LHII
(the first four chromophores from the $N=18$ example of the MC-VQE theory paper,
with nearest neighbor linear connectivity and dipole-dipole couplings). In all
cases, we prepare a number $N_{\Theta}$ of ``contracted reference states,'' $|
\Phi_{\Xi} \rangle \equiv \sum_{\vec I} C_{\vec I \Xi} | \vec I \rangle$ taken
here from classical configuration interaction singles (CIS) with restricted
configurations $| \vec I \rangle$ including the reference configuration
$|0000\rangle$ and all singly-excited configurations, e.g.,
$|0010\rangle$. A simple quantum circuit to prepare such states was shown
in our MC-VQE paper. These are then correlated with a VQE entangler circuit
$\hat U (\{ \theta_g \})$ so as to minimize the expectation value of the
state-averaged energy,
\begin{equation}
\mathcal{O} (\{ \theta_g \})
\equiv
\frac{1}{N_{\Theta}}
\sum_{\Xi}^{N_{\Theta}}
\langle \Phi_{\Xi} |
\hat U^{\dagger} (\{ \theta_g \})
\hat H
\hat U (\{ \theta_g \})
| \Phi_{\Xi} \rangle
\end{equation}
Subsequent to state-averaged VQE optimization, the contracted subspace
Hamiltonian $H_{\Xi \Xi'} \equiv \langle \Phi_{\Xi} | \hat U^\dagger (\{
\theta_g \} )\hat H \hat U (\{ \theta_g \}) | \Phi_{\Xi} \rangle$ can be
computed by additional quantum measurements, then classically diagonalized to
form the adiabatic MC-VQE states $| \Psi_{\Theta} \rangle$ and corresponding
diagonal or transition properties. The selected state-averaged VQE entangler
circuit is,
\begin{widetext}
\begin{equation}
\Qcircuit @R=1.0em @C=0.5em {
\lstick{|A\rangle}
 & \gate{R_{y} (\theta_{1})}
 & \ctrl{1}
 & \gate{R_{y} (\theta_{5})}
 & \ctrl{1}
 & \gate{R_{y} (\theta_{9})}
 & \qw & \qw & \qw & \qw
 & \qw \\
\lstick{|B\rangle}
 & \gate{R_{y} (\theta_{2})}
 & \targ
 & \gate{R_{y} (\theta_{6})}
 & \targ
 & \gate{R_{y} (\theta_{10})}
 & \ctrl{1}
 & \gate{R_{y} (\theta_{13})}
 & \ctrl{1}
 & \gate{R_{y} (\theta_{15})}
 & \qw \\
\lstick{|C\rangle}
 & \gate{R_{y} (\theta_{3})}
 & \ctrl{1}
 & \gate{R_{y} (\theta_{7})}
 & \ctrl{1}
 & \gate{R_{y} (\theta_{11})}
 & \targ
 & \gate{R_{y} (\theta_{14})}
 & \targ
 & \gate{R_{y} (\theta_{16})}
 & \qw \\
\lstick{|D\rangle}
 & \gate{R_{y} (\theta_{4})}
 & \targ
 & \gate{R_{y} (\theta_{8})}
 & \targ
 & \gate{R_{y} (\theta_{12})}
 & \qw & \qw & \qw & \qw 
 & \qw \\
}
\end{equation}
\end{widetext}
Note the lexical ordering of $N_g = 16$ gate angles, arbitrarily chosen to run
with the qubit index fast and the time index slow. 

The origins of the ``easy'' and ``hard'' test cases come from the consideration
of the dense spectrum of the excited states $| \Psi_1 \rangle$ to $|\Psi_5
\rangle$ - under the exact full configuration interaction (FCI) solution, these
lie between 1.84 and 2.10 eV above the ground state, and have markedly different
oscillator strengths.  The ``easy'' test case is chosen by selecting $N_\Theta =
5$ - here, the specifics of the CIS contracted reference states are not
important, as the whole singles manifold is covered. In this case, the goal of
the state-averaged VQE entangler circuit is to decouple the ground and first
four excited states (dominated by linear combinations of single excitations)
from the rest of the Hilbert space - the subsequent subspace diagonalization
procedure will handle rotations within the entangled contracted reference states
to form the final MC-VQE adiabatic states. The ``hard'' test case is chosen by
selecting $N_\Theta = 3$ - here, the state-averaged VQE entangler circuit must
also decouple the target states from states $|\Psi_{4} \rangle$ and $|\Psi_{5}
\rangle$, involving considerable singles-singles mixing. This is analogous to
targeting a subset of densely packed eigenstates in classical linear algebra
algorithms, which is known to cause slow convergence in methods such as Lanczos,
or Davidson-Liu. In practice, we find that the ``hard'' case is generally much
more difficult to converge to any given local minimum than the ``easy'' case,
exhibits more significantly multiple local minima than we observe in the
``easy'' case, and has higher errors for absolute and difference energies and
transition properties than the ``easy'' case, regardless of which local minimum
is converged. 

In the numerical tests, we compare the standard L-BFGS and Powell
implementations in \textsc{SciPy} to various Jacobi methods. For the Jacobi
methods, we consider the standard ``Jacobi-1'' [all single angle pivots
$(\theta_g,)$, in lexical order] and ``Jacobi-2'' [all double angle pivots
$(\theta_g, \theta_g') : \ g \geq g'$]. We also explore the use of the
generalized Jacobi recipe by considering two variants of sieved Jacobi-2:
``Jacobi-A,'' in which all two-angle pivots on the same qubit index are included
[e.g., $(\theta_5, \theta_1)$, $(\theta_9, \theta_1)$, and $(\theta_9,
\theta_5)$, but not $(\theta_2, \theta_1)$ or $(\theta_3,\theta_1)$], and
``Jacobi-B,'' in which all two-angle pivots on linearly adjacent qubit indices
are included [e.g., adding $(\theta_2, \theta_1)$ but not $(\theta_3, \theta_1)$
to the Jacobi-A example].  For this example, Jacobi-1 requires 3-point
tomography of 16 gates, for 48 total observables per iteration (reducible to 33
observables per iteration if tomography information is reused between gate
optimizations). Jacobi-2 requires 9-point tomography of 120 gate pairs, for 1080
total observables per iteration.  Jacobi-A requires 9-point tomography of 26
gate pairs, for 234 total observables per iteration. Jacobi-B requires 9-point
tomography of 81 gate pairs, for 729 total observables per iteration.
Gradient-based methods like L-BFGS require 33 at least observables per
iteration: for L-BFGS we find typically $\sim36$ observables per iteration, with
line searches to updated trust regions. Powell requires a line search per gate
per iteration: we find tyically $\sim 220$ observables per iteration.

We also explore various convergence acceleration approaches. One possibility is
to randomize the pivot order of the standard Jacobi procedure, a method denoted
``Jacobi-$F$-Random.'' We also use Anderson acceleration applied with the
traditional Anderson error vector of the move length $\{ \delta
\theta_{g}^{\prime} \}^{k}$ (``Jacobi-$F$-Anderson'') and with the Pulay DIIS
error vector of the energy gradient $ \{ \pdiff{\mathcal{O}}{\theta_g} \}^{k}$
(``Jacobi-$F$-Pulay). The second choice additionally requires the computation of
the gradient after each Jacobi sweep, bringing the total number of observables
per iteration to 81. For the DIIS procedure, we store at most 10 iterative
history vectors with a worst-error removal policy, and we flush the DIIS
iterative history every 40 iterations. We have tested mixing pivot randomization
and Anderson/Pulay variants of DIIS, and have found (as expected) that the DIIS
procedure cannot tolerate the randomized pivots and does not provide any
improvement over Jacobi-Random. Therefore, we do not explicitly discuss mixings
of Jacobi-Rand and Jacobi-Anderson/Pulay below.

In all cases, we start from a guess of zero entanglement $\{ \theta_g = 0 \}$.
We converge each algorithm for 100 iterations or until a maximum gradient
element in the state-averaged energy falls below $1\times10^{-7}$.

\section{Results and Discussion}

Including the classes of Jacobi-1, Jacobi-2, Jacobi-A, and Jacobi-B and
their direct product with the choices of no convergence acceleration, pivot randomization,
Anderson, and Pulay, plus the reference L-BFGS and Powell methods, there are $18
\times$ optimization methods to test. To proceed, we will compare the
convergence of different methods within a given Jacobi class (wherein logical
iterations for the various convergence acceleration methods all cost the same
number of observable expectations values, up to the additional modest gradient
evaluation cost of Pulay), to downselect the convergence acceleration method to
one case per Jacobi class.  We will then compare the different selected Jacobi
methods to each other and to L-BFGS and Powell first with respect to logical
iteration count, and finally with respect to convergence as a function of number
of observable expectation values. The latter is the ultimate cost function:
optimization of variational quantum circuits within the statistical and device
noise limit on near-term quantum hardware requires optimization algorithms with
rapid initial convergence relative to the number of required observable
expectation values, while classical benchmark testing requires optimization
algorithms with fast and robust global convergence relative to the number of
required observable expectation values.

\subsection{Jacobi-1 Methods (All Single Angles)}

\begin{figure}[h!]
\begin{center}
\includegraphics[width=3.4in]{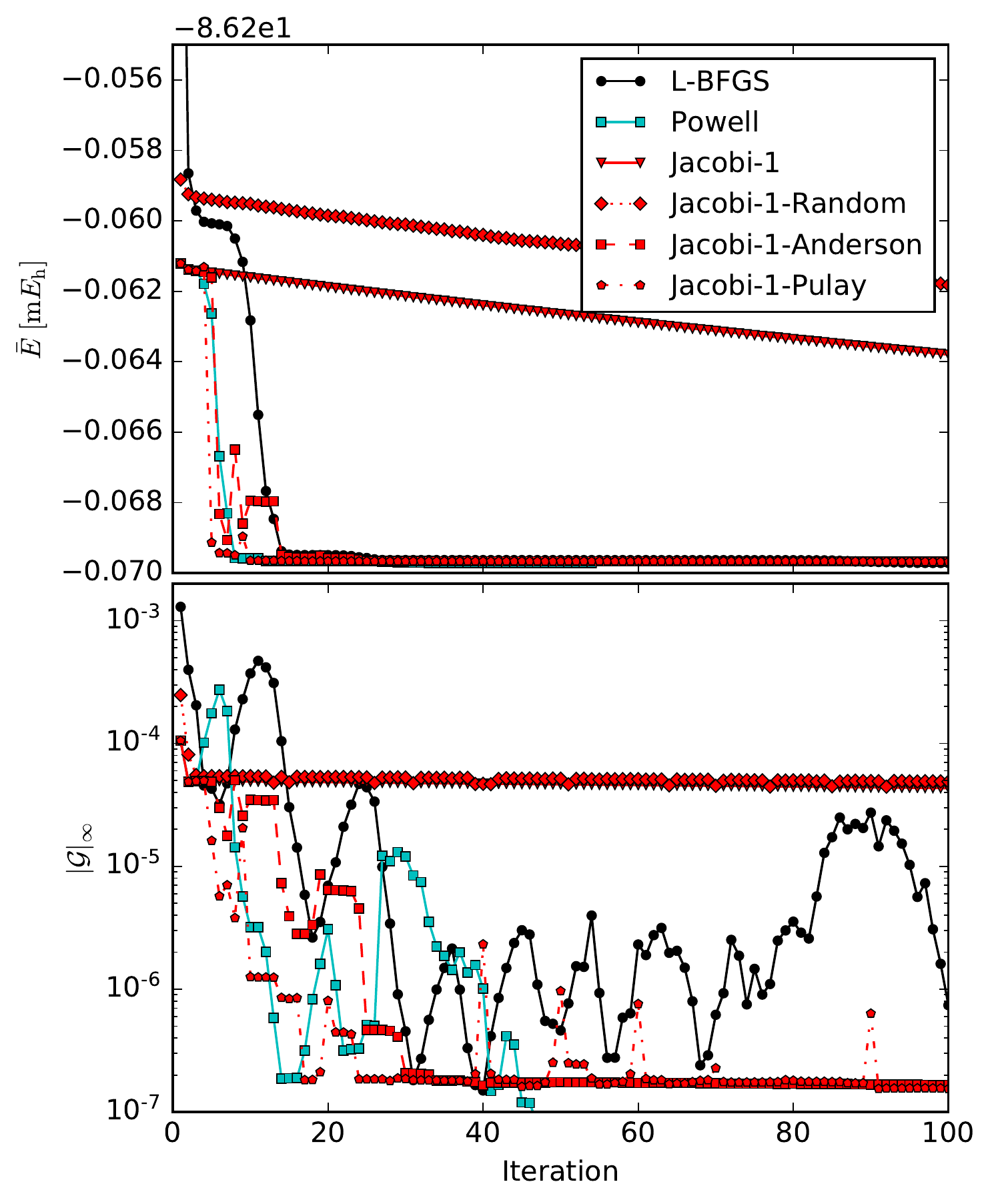}
\caption{State-averaged VQE energy convergence (top) and maximum gradient
element (bottom) as a function of logical iteration count for the $N_{\mathrm{state}}=5$ ``easy''
test case. Jacobi-1 (all single angles) methods highlighted.}
\label{fig:j1-5B}
\end{center}
\end{figure}

Figure \ref{fig:j1-5B} shows the convergence of the Jacobi-1 methods for the
$N_{\mathrm{state}}$ ``easy'' test case. The first finding is that a single
iteration of Jacobi-1 can provide a relatively good starting point that is
substantially lower in energy and maximum gradient than the result of the first
logical L-BFGS iteration (not visible in the energy figure), and identical in
quality to the first logical Powell step. The latter observation is expected:
the first step of Jacobi-1 and Powell are identical in spirit, though the
Jacobi-1 recipe has an analytical 3-point formula that yields and exact
linesearch. The second finding is that the unaccelerated Jacobi-1 methods
experience rapid convergence stagnation for these types of variational quantum
circuits. This is indicated by the slow monotonic decay of the energy for
Jacobi-1 and Jacobi-1-Random. A third finding is that pivot randomization seems
to have substantial effect on the quality of the first iteration, but does not
affect the convergence stagnation (emphasis that the pivots are randomized in
every logical iteration). This indicates that other choices of deterministic
Jacobi-1 pivot order should be explored to achieve the best initial iteration,
but also indicates that pivot randomization should not be pursued. A fourth
finding is that Jacobi-1-Anderson, Jacobi-1-Pulay, and Powell all perform
remarkably similarly during the first part of the iterative procedure: all
rapidly settle near a fixed point that appears to be associated with a saddle
point or local minimum for a few iterations, and then all turn over at iteration
4 or 5 and converge rapidly toward the presumed global minimum. This indicates
that the Anderson or Pulay convergence acceleration procedures can be successful
at analyzing the iterative histories of moves and moving ``diagonally'' across
the parameter space to target deeper minima and enhance convergence. The
correspondence of this turnover with that of Powell is striking, particularly
because Jacobi-1-Anderson/Pulay and Powell are operating by different mechanisms
(and the Jacobi-1-Anderson/Pulay methods are several times cheaper than Powell
due to the ability to forgo a linesearch for each search direction). The final
finding is that the Anderson/Powell iterations do exhibit some oscillation in
the terminal convergence epoch (particularly as the DIIS iterative history is
flushed every 10 iterations), and end up stagnating at a maximum gradient
element of $\sim 10^{-7}$. Powell seems to be more robust for extremely tight
convergence - to mitigate this behavior, we can either develop methods to dampen
the DIIS flushes, add characteristics to the DIIS iterative history to penalize
for increased energy during the iterations as is done in the EDIIS
extension,\cite{kudin2002black} or simply switch to Powell in the later stages
of very tight classical optimizations needed for benchmarking variational
quantum circuits.

\begin{figure}[h!]
\begin{center}
\includegraphics[width=3.4in]{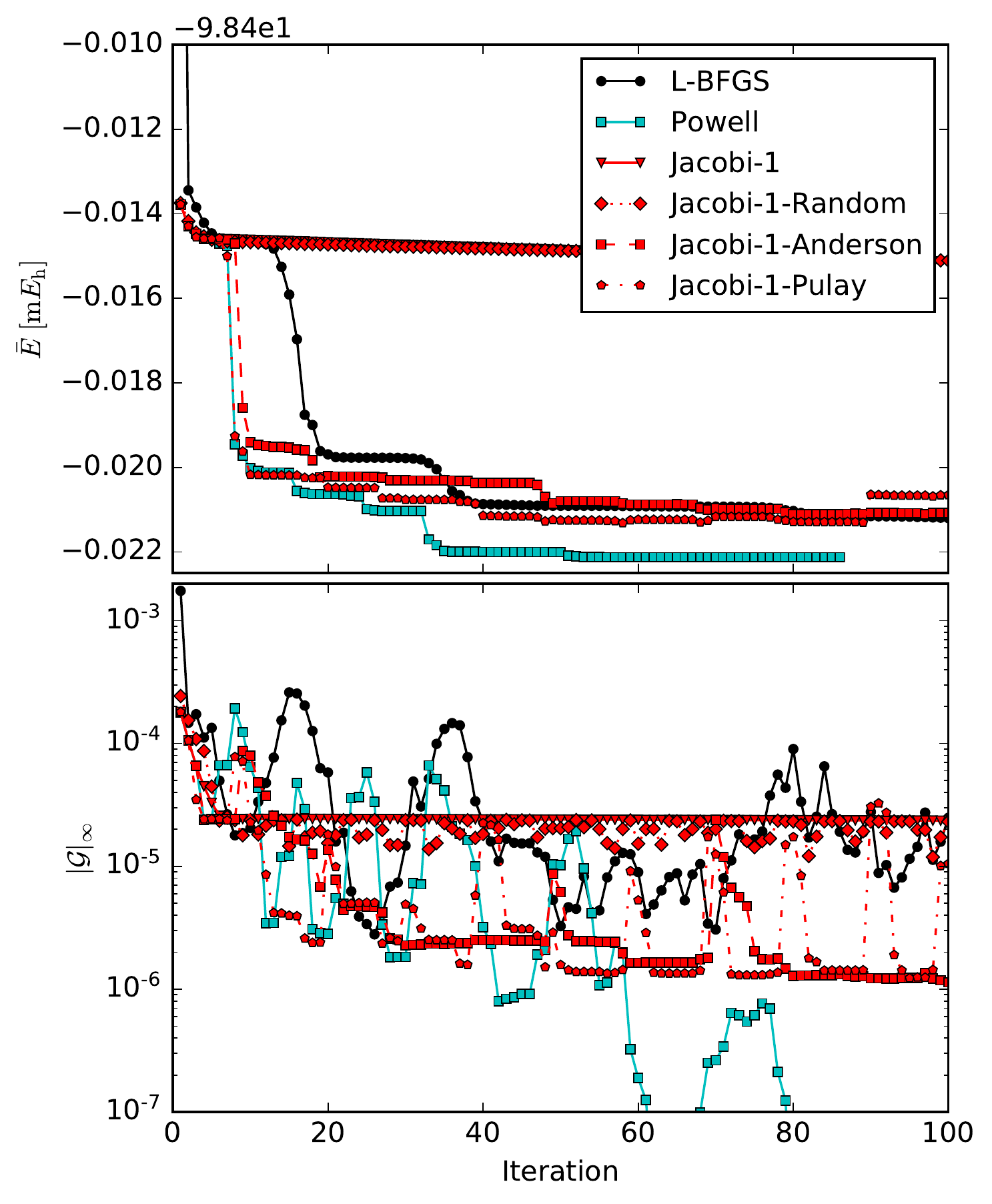}
\caption{State-averaged VQE energy convergence (top) and maximum gradient
element (bottom) as a function of logical iteration count for the $N_{\mathrm{state}}=3$ ``hard''
test case. Jacobi-1 (all single angles) methods highlighted.}
\label{fig:j1-3B}
\end{center}
\end{figure}

Figure \ref{fig:j1-3B} shows the convergence of the Jacobi-1 methods for the
$N_{\mathrm{state}}$ ``hard'' test case. The findings for this case are broadly
similar as for the ``easy'' test case: the rapid early start relative to L-BFGS,
expected stagnation of Jacobi-1 and Jacobi-1-Random,
and remarkable correspondence of Jacobi-1-Anderson/Pulay to Powell continue. The
primary difference in this case is the appearance of multiple stagnation points
in the more-complicated objective function landscape for the hard problem.
Though the Jacobi-1-Anderson/Pulay methods initially follow the Powell method,
in later methods they slow down and more-closely follow the L-BFGS convergence.
Powell remains the most robust method shown for tight convergence, and produces
a markedly lower energy than predicted by any Jacobi-1 or L-BFGS method for this
hard test case, albeit with several times more objective function evaluations
per logical iteration.

For the global comparison below, we select the Jacobi-1-Pulay method for
the Jacobi-1 class, as it exhibits fewer/smaller oscillations near DIIS
restarts. However, we reiterate that improving the overall monotonicity of the
DIIS procedure is certainly a worthy topic for future study.

\subsection{Jacobi-2 Methods (All Pairs of Angles)}

\begin{figure}[h!]
\begin{center}
\includegraphics[width=3.4in]{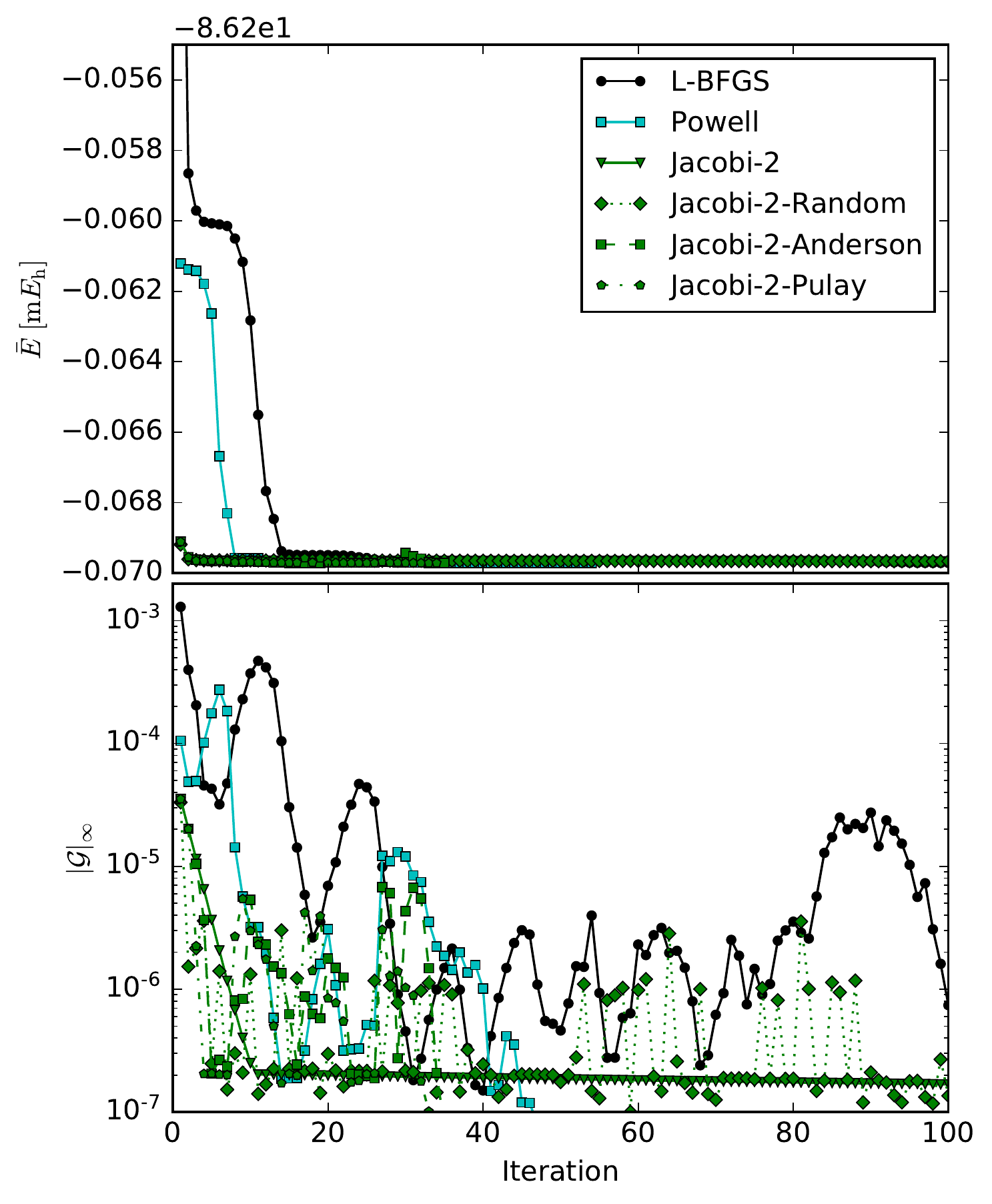}
\caption{State-averaged VQE energy convergence (top) and maximum gradient
element (bottom) as a function of logical iteration count for the $N_{\mathrm{state}}=5$ ``easy''
test case. Jacobi-2 (all pairs of angles) methods highlighted.}
\label{fig:j2-5B}
\end{center}
\end{figure}

Figure \ref{fig:j2-5B} shows the convergence of the Jacobi-2 methods for the
$N_{\mathrm{state}}$ ``easy'' test case.  The first finding is that the result
after the first logical iteration of Jacobi-2 is substantially lower in energy
and maximum gradient element than for Powell or the previously considered
Jacobi-1, albeit at substantially higher cost in terms of require number of
observable expectation values. The second finding is that pivot randomization
does not improve convergence, as expected. The third finding is that neither
Anderson nor Pulay convergence acceleration appear to offer significant
improvement over the natural fixed-point convergence of Jacobi-2. This is not
unexpected given the strong relationship of DIIS to second-order quasi-Newton
methods, together with the fact that Jacobi-2 is using information equivalent to
the full Hessian - it is likely that the natural second-order aspect of Jacobi-2
is redundant with the approximate second-order DIIS procedure. The last finding
is that the convergence of Jacobi-2 stagnates at a maximum gradient element of
$\sim 10^{-2}$, while the Powell method is immune to this problem - as before,
Powell is superior for very tightly converged optimizations.

\begin{figure}[h!]
\begin{center}
\includegraphics[width=3.4in]{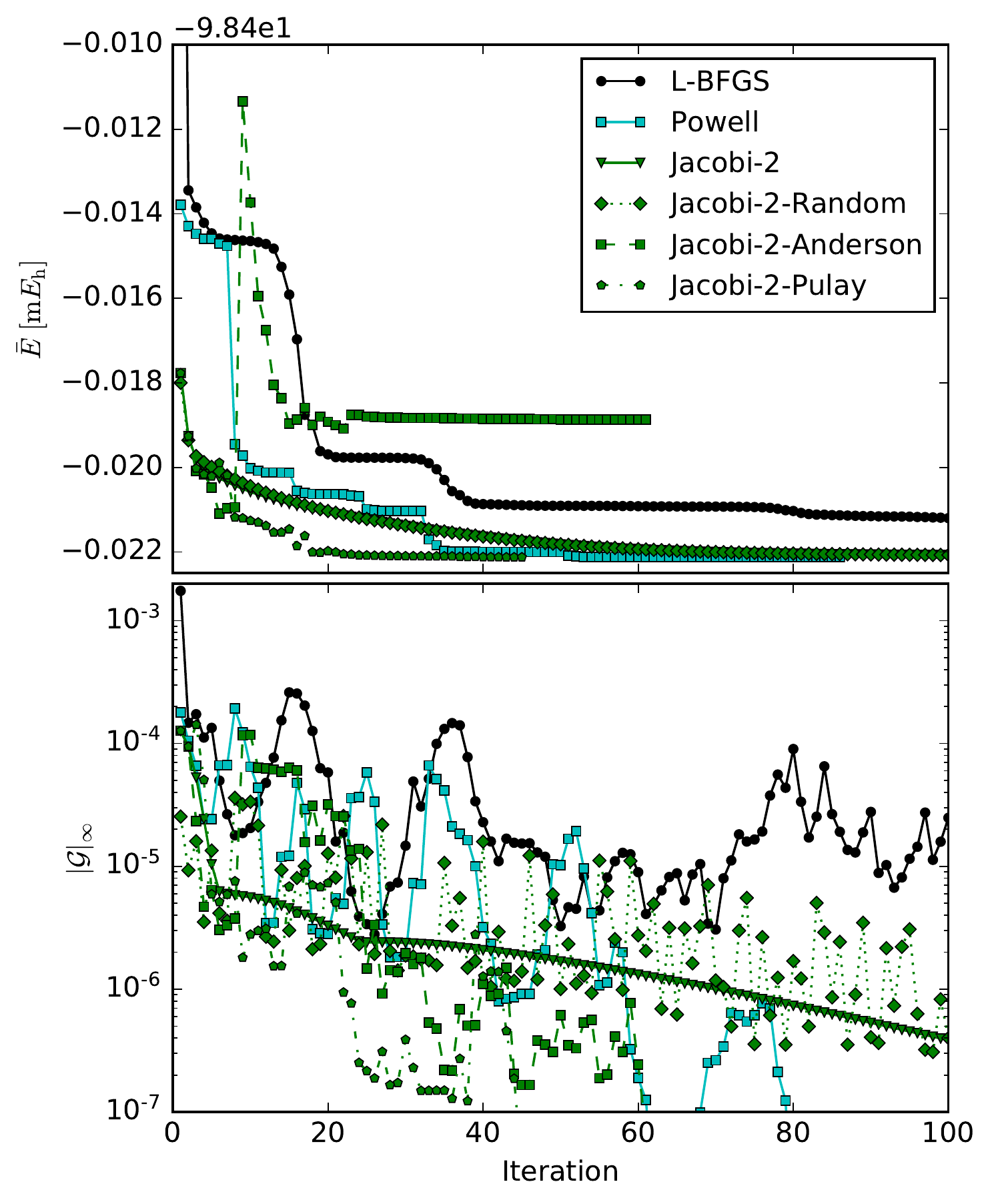}
\caption{State-averaged VQE energy convergence (top) and maximum gradient
element (bottom) as a function of logical iteration count for the $N_{\mathrm{state}}=3$ ``hard''
test case. Jacobi-2 methods (all pairs of angles) highlighted.}
\label{fig:j2-3B}
\end{center}
\end{figure}

Figure \ref{fig:j2-3B} shows the convergence of the Jacobi-2 methods for the
$N_{\mathrm{state}}$ ``hard'' test case.  The results are similar to the easy
case, with a few caveats. The first is that the starting point and convergence
rate of Jacobi-1 are somewhat worse than the easy case. The second is that while
the Pulay variant of DIIS seems to provide some improvement in energy
convergence of Jacobi-2, the Jacobi-2-Anderson method exhibits an energy
oscillation during the DIIS flush procedure at the 10th iteration. and then
converges to a different, higher local minimum.

For the global comparison below, we select the Jacobi-2 method for the
Jacobi-2 class, due to the apparent redundancy between the second-order aspects
of Jacobi-2 and the approximate second-order DIIS method.

\subsection{Jacobi-A Methods (All Pairs of Angles Within Single Qubit Wires)}

\begin{figure}[h!]
\begin{center}
\includegraphics[width=3.4in]{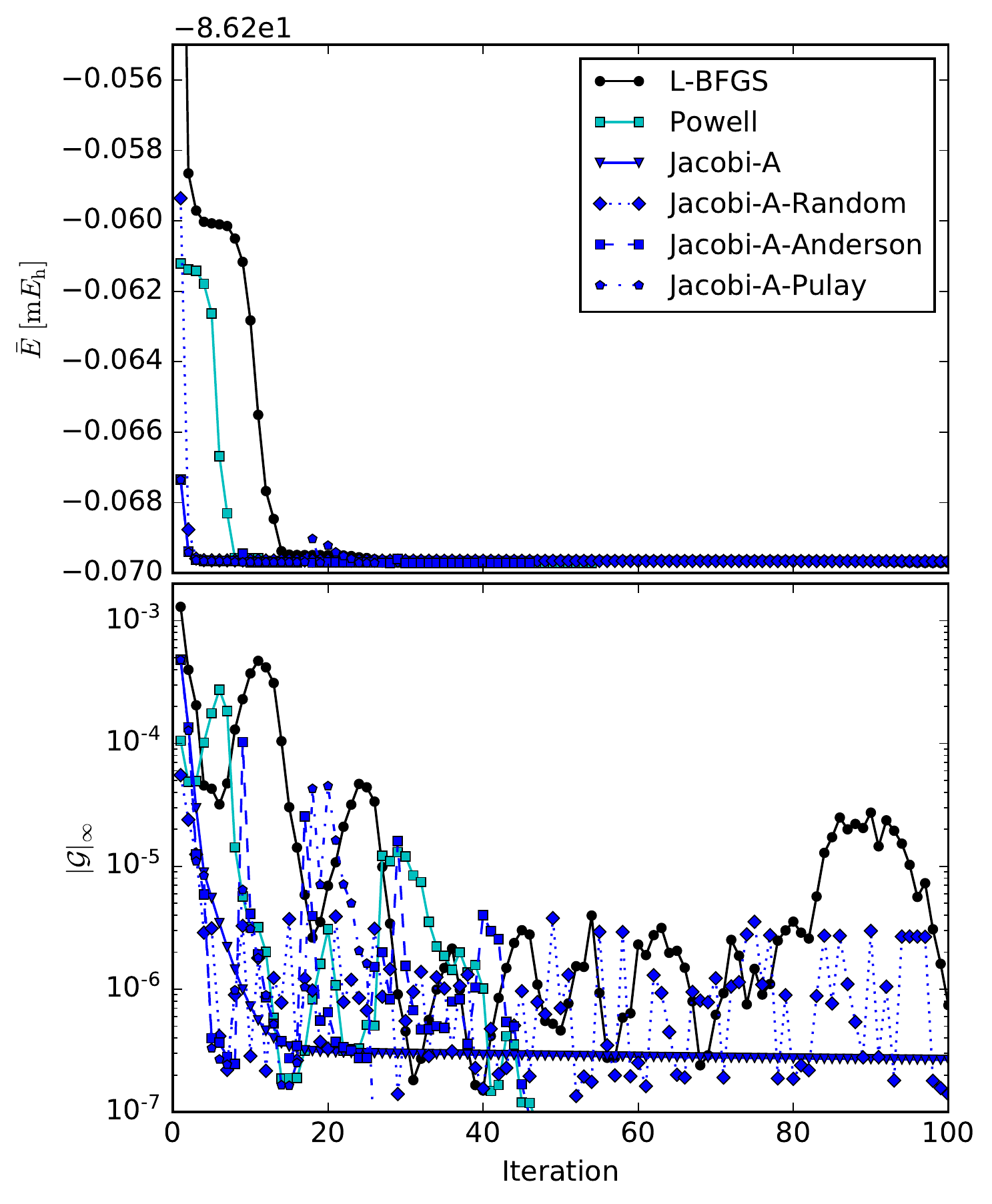}
\caption{State-averaged VQE energy convergence (top) and maximum gradient
element (bottom) as a function of logical iteration count for the $N_{\mathrm{state}}=5$ ``easy''
test case. Jacobi-Gen ``A'' methods (all pairs of angles within single qubit
wires) highlighted.}
\label{fig:jA-5B}
\end{center}
\end{figure}

\begin{figure}[h!]
\begin{center}
\includegraphics[width=3.4in]{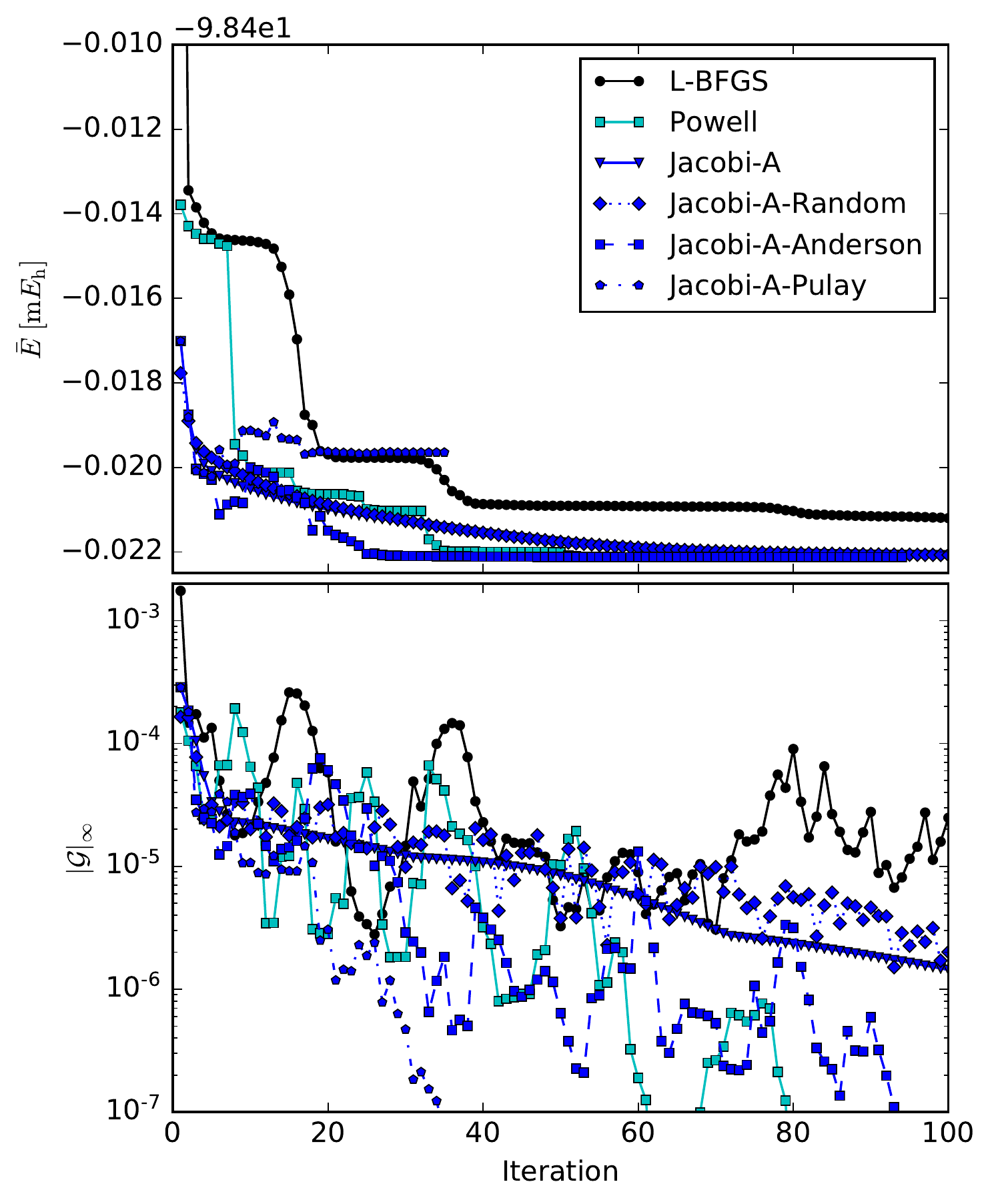}
\caption{State-averaged VQE energy convergence (top) and maximum gradient
element (bottom) as a function of logical iteration count for the $N_{\mathrm{state}}=3$ ``hard''
test case. Jacobi-Gen ``A'' methods (all pairs of angles within single qubit
wires) highlighted.}
\label{fig:jA-3B}
\end{center}
\end{figure}

Figures \ref{fig:jA-5B} and \ref{fig:jA-3B} show the convergence of the Jacobi-A
methods for the $N_{\mathrm{state}}$ ``easy'' and ``hard'' test cases,
respectively. The results from the Jacobi-A methods are remarkably similar to
those from the Jacobi-2 methods, despite the angle pairs being restricted to be
within individual qubit wires for Jacobi-A. Also worth noting is that the
Jacobi-A-Pulay method exhibits a similar energy oscillation during the DIIS
flush procedure and convergence to a different local minimum, similar as was
seen for this test case in Jacobi-2-Anderson. This indicates that the energy
oscillations and subsequently problematic convergence with DIIS acceleration in
Jacobi-2 methods is not specific to the choice of Anderson or Pulay error
vectors. As with Jacobi-2, we select the unaccelerated Jacobi-A method for the
global comparison below.

\subsection{Jacobi-B Methods (All Pairs of Angles Within Single and On Adjacent
Qubit Wires)}

\begin{figure}[h!]
\begin{center}
\includegraphics[width=3.4in]{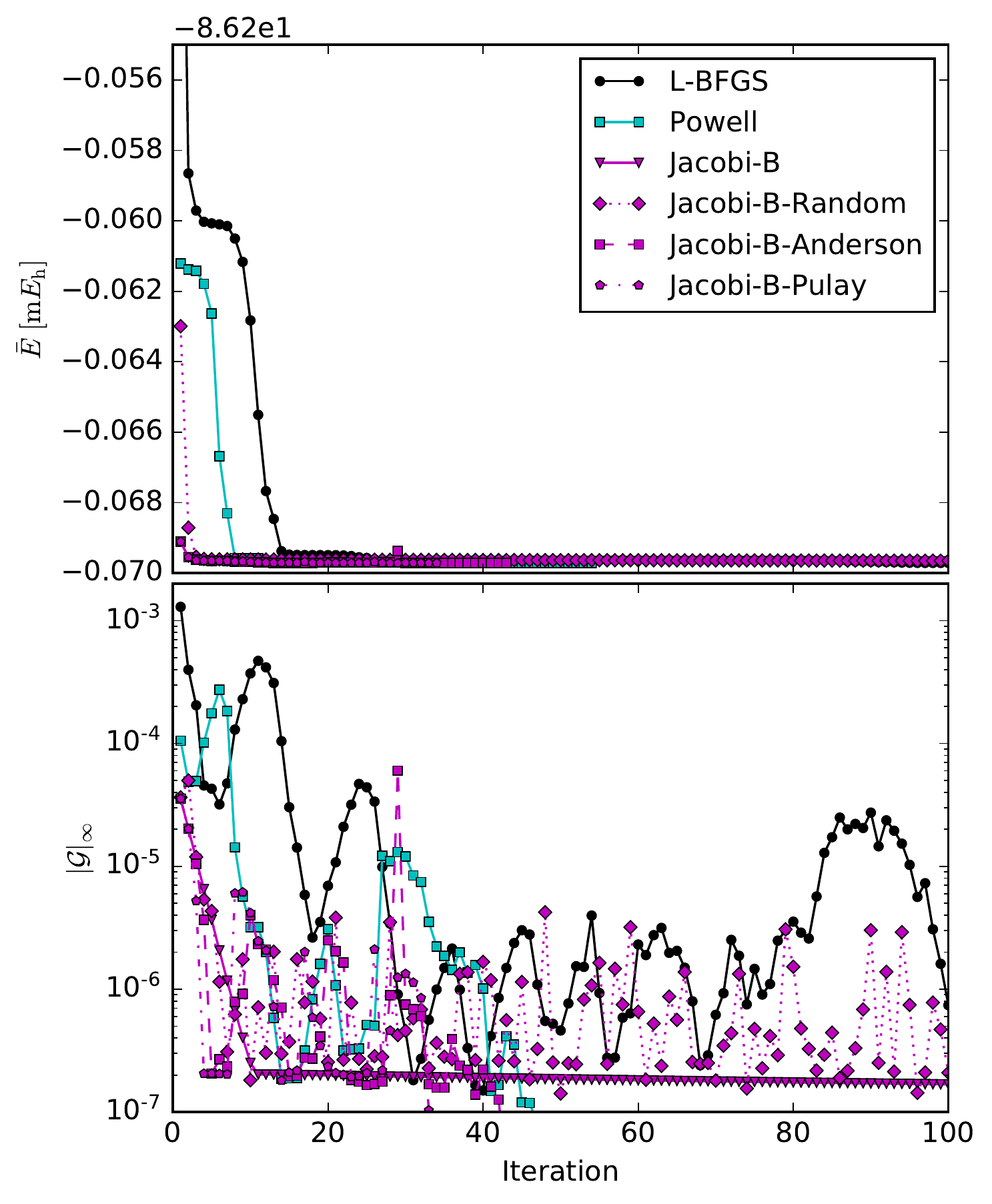}
\caption{State-averaged VQE energy convergence (top) and maximum gradient
element (bottom) as a function of logical iteration count for the $N_{\mathrm{state}}=5$ ``easy''
test case. Jacobi-Gen ``B'' methods (all pairs of angles within single and on
adjacent qubit wires) highlighted.}
\label{fig:jB-5B}
\end{center}
\end{figure}

\begin{figure}[h!]
\begin{center}
\includegraphics[width=3.4in]{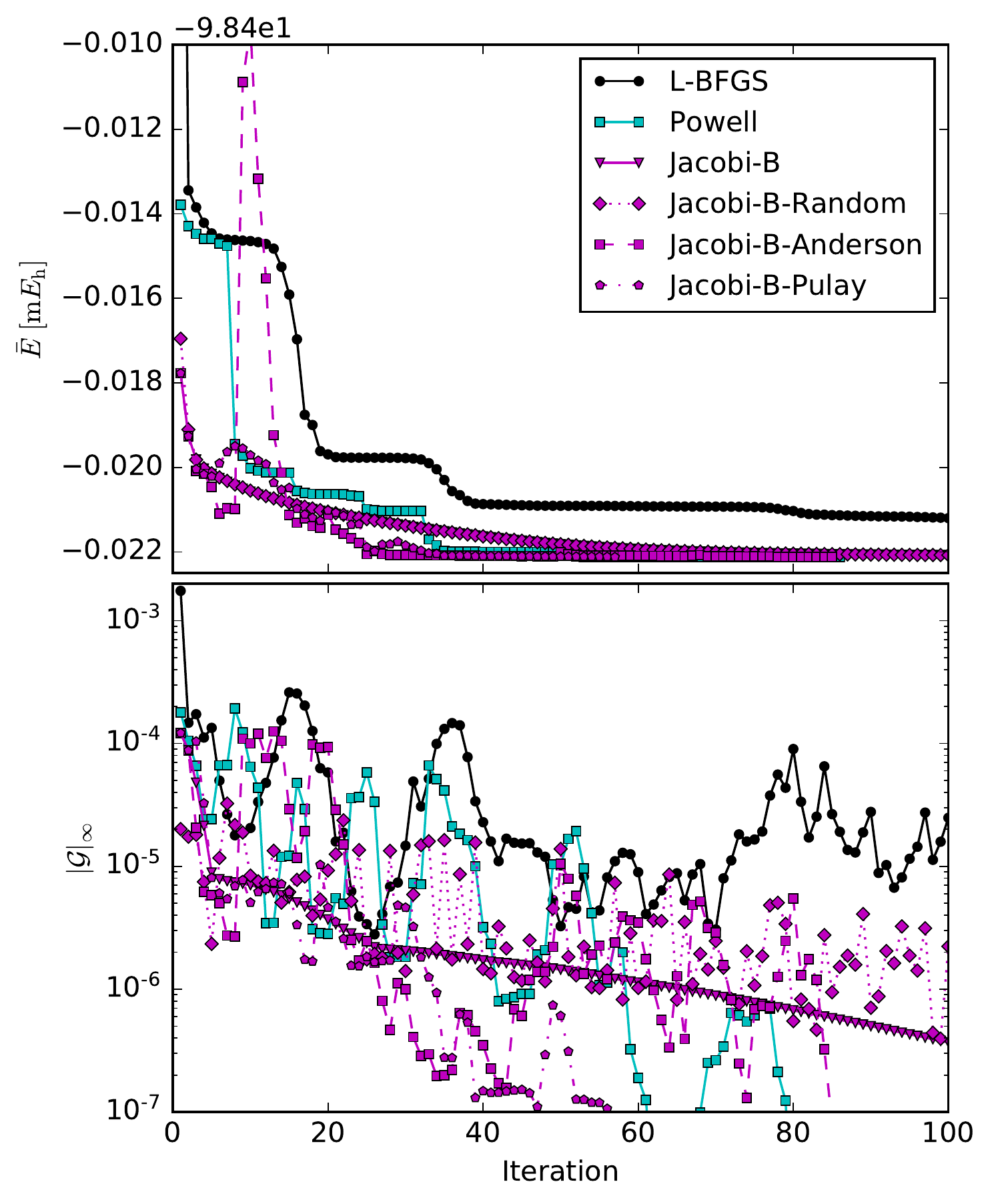}
\caption{State-averaged VQE energy convergence (top) and maximum gradient
element (bottom) as a function of logical iteration count for the $N_{\mathrm{state}}=3$ ``hard''
test case. Jacobi-Gen ``B'' methods (all pairs of angles within single and on
adjacent qubit wires) highlighted.}
\label{fig:jB-3B}
\end{center}
\end{figure}

Figures \ref{fig:jB-5B} and \ref{fig:jB-3B} show the convergence of the Jacobi-A
methods for the $N_{\mathrm{state}} = 5$ and $3$, ``easy'' and ``hard'' test
cases, respectively. The results are entirely congruent with the findings for
the Jacobi-2 and Jacobi-A methods above, which is not surprising as the angle
pair constituents of Jacobi-B fall between Jacobi-2 and Jacobi-A. As with
Jacobi-2 or Jacobi-A, we select the unaccelerated Jacobi-B method for the global
comparison below.

\subsection{Global Optimization Method Comparison vs. Logical Iteration Number}

\begin{figure}[h!]
\begin{center}
\includegraphics[width=3.4in]{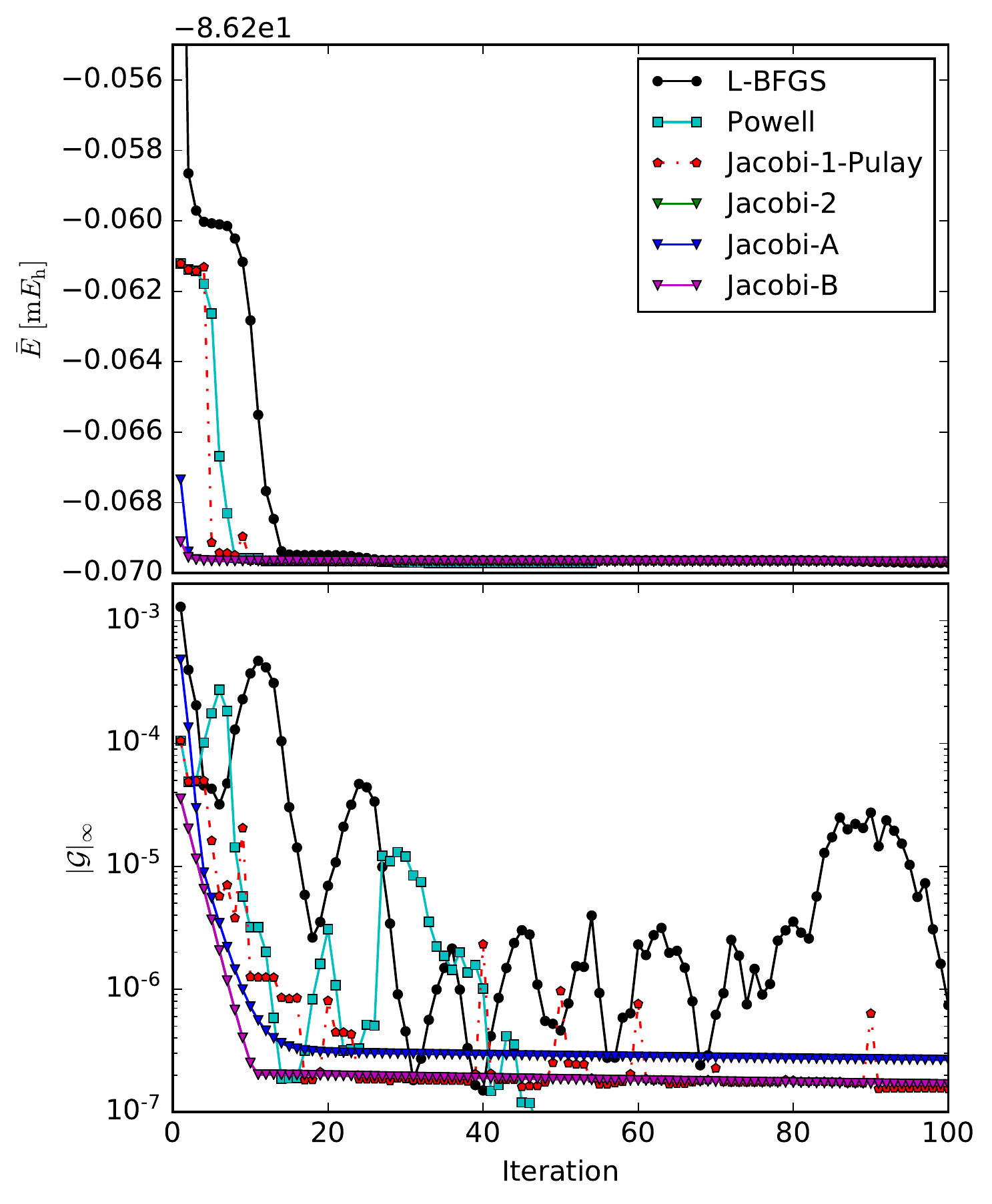}
\caption{State-averaged VQE energy convergence (top) and maximum gradient
element (bottom) as a function of logical iteration count for the $N_{\mathrm{state}}=5$ ``easy''
test case. Jacobi-1-Pulay, Jacobi-2, Jacobi-A, and Jacobi-B methods compared to
L-BFGS and Powell.}
\label{fig:final-5B}
\end{center}
\end{figure}

Figure \ref{fig:final-5B} shows the convergence of the selected Jacobi methods and
the L-BFGS and Powerll methods for the $N_{\mathrm{state}} = 5$ ``easy'' test case.  
The results show that Jacobi-1-Pulay and Powell have a highly similar
convergence history, thought the Jacobi-1-Pulay method is several times cheaper.
Both Jacobi-1-Pulay and Powell exhibit faster logical convergence than L-BFGS.
Jacobi-A, Jacobi-B, and Jacobi-2 all exhibit even more improved initial
iterations, and converge logically faster than the other methods, but cost
considerably more observable expectation values per iteration. The Jacobi
methods all stagnate at a maximum gradient element of $\sim 10^{-7}$, while the
Powell method eventually drops below this value - therefore Powell's method
remains superior for extremely tightly converged optimizations. A final note is
the correspondence between Jacobi-A/B/2: all three are similar, while Jacobi-B
and Jacobi-2 are essentially coincident.

\begin{figure}[h!]
\begin{center}
\includegraphics[width=3.4in]{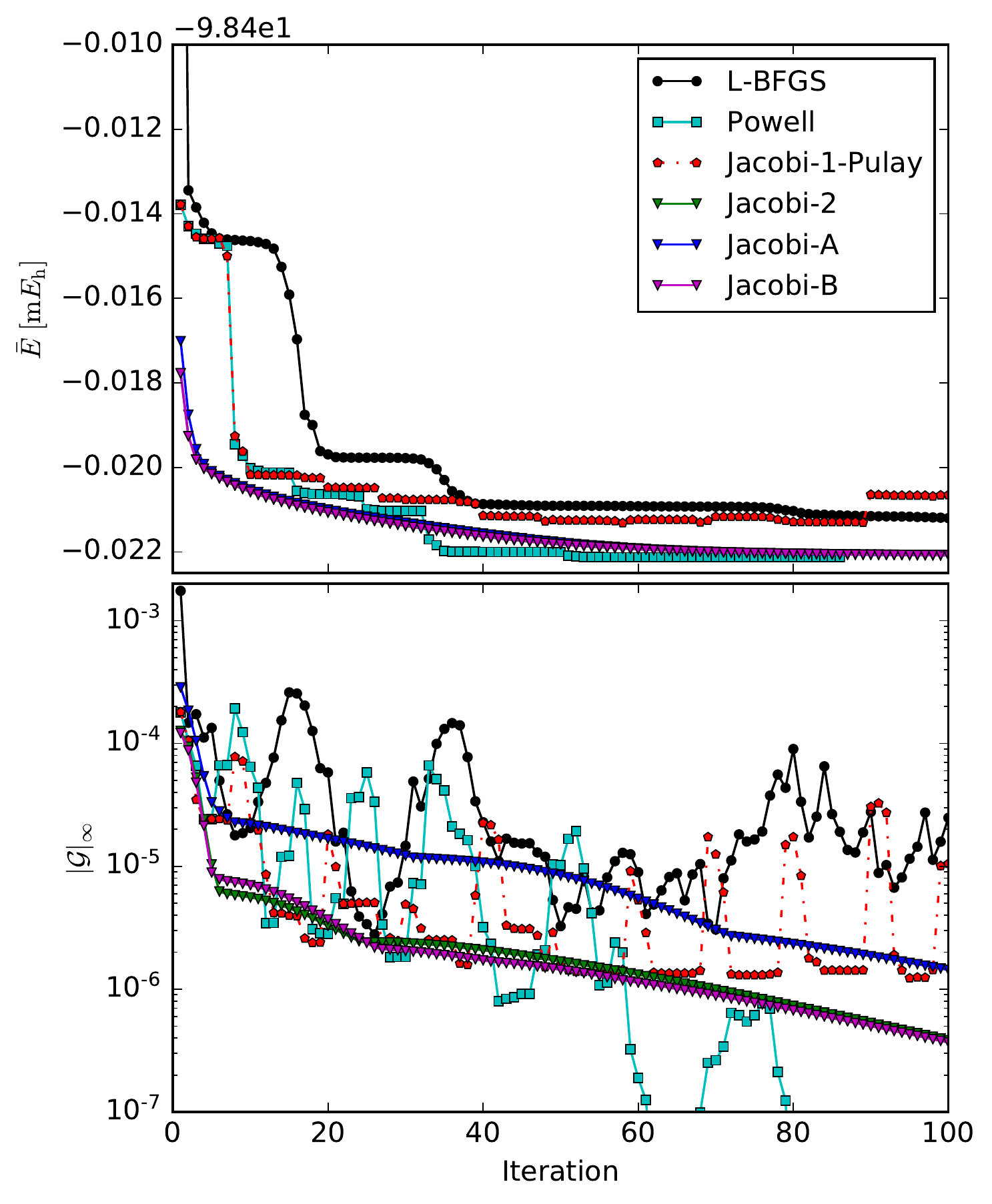}
\caption{State-averaged VQE energy convergence (top) and maximum gradient
element (bottom) as a function of logical iteration count for the $N_{\mathrm{state}}=3$ ``hard''
test case. Jacobi-1-Pulay, Jacobi-2, Jacobi-A, and Jacobi-B methods compared to
L-BFGS and Powell.}
\label{fig:final-3B}
\end{center}
\end{figure}

Figure \ref{fig:final-3B} shows the convergence of the selected Jacobi methods
and the L-BFGS and Powell methods for the $N_{\mathrm{state}} = 3$ ``hard'' test
case. The results are similar to those for the ``easy'' test case, with the
exception that the starting points and convergence rates are somewhat worse for
all methods, and that the Jacobi-1-Pulay method starts out coincident with the
logical iteration progress of Powell, but eventually degrades to be more
coincident with L-BFGS.

\subsection{Global Optimization Method Comparison vs. Observable Expectation
Value Count}

\begin{figure}[h!]
\begin{center}
\includegraphics[width=3.4in]{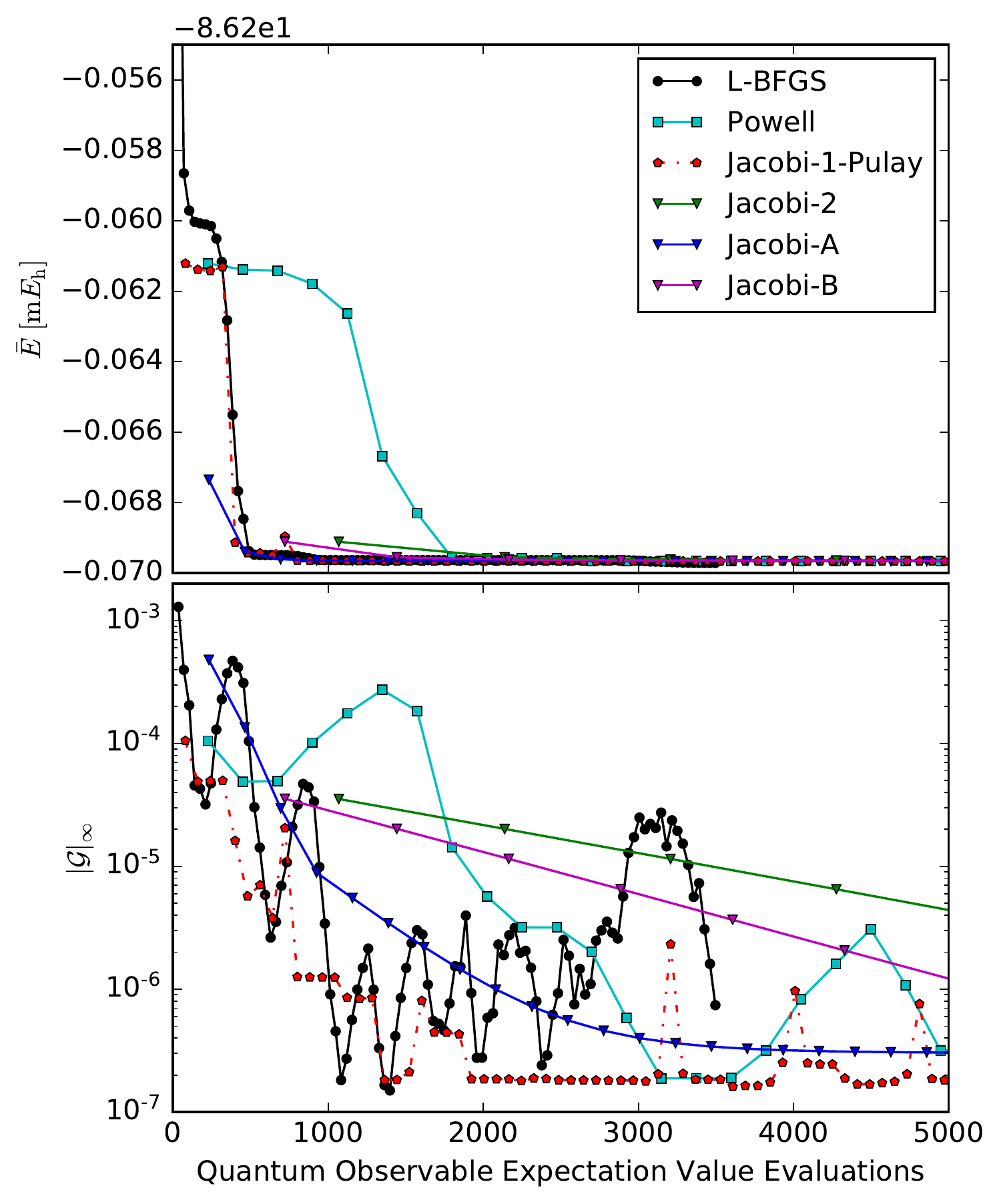}
\caption{State-averaged VQE energy convergence (top) and maximum gradient
element (bottom) as a function of observable expectation value count for the
$N_{\mathrm{state}}=5$ ``easy'' test case. Jacobi-1-Pulay, Jacobi-2, Jacobi-A, and Jacobi-B
methods compared to L-BFGS and Powell.}
\label{fig:final-5B2}
\end{center}
\end{figure}

For deployment of variational quantum algorithms on near-term hardware, a key
challenge is to minimize the number of observables expectation values that must
be statistically resolved to optimize the circuit parameters. In Figure
\ref{fig:final-5B2}, we compare the early convergence history of the selected
Jacobi methods with L-BFGS and Powell for the ``easy'' test case. Here the
overhead of bidirectional linesearch in Powell becomes apparent: the early
convergence history of Jacobi-1-Pulay and Power are highly similar, but the
Powell curve is scaled by a factor of $\sim 2.6$. Jacobi-1-Pulay achieves early
gains over L-BFGS, but becomes nearly coincident with L-BFGS as the optimization
progresses. Note that there would be an additional gain if Jacobi-1-Anderson
were plotted here due to the reduced overhead from the computation of the
gradient needed for the Jacobi-1-Pulay error vector, but we did not select
Jacobi-1-Anderson due to oscillations present in the DIIS flush procedure in
later iterations. The Jacobi-A method is also notable: its first iteration
achieves a high accurate-to-cost-ratio starting point that is well below
the false minimum seen in the first few iterations of Jacobi-1-Pulay and Powell.
Variations on Jacobi-A might prove useful to provide robust but low cost initial
starting points near the global minimum.

\begin{figure}[h!]
\begin{center}
\includegraphics[width=3.4in]{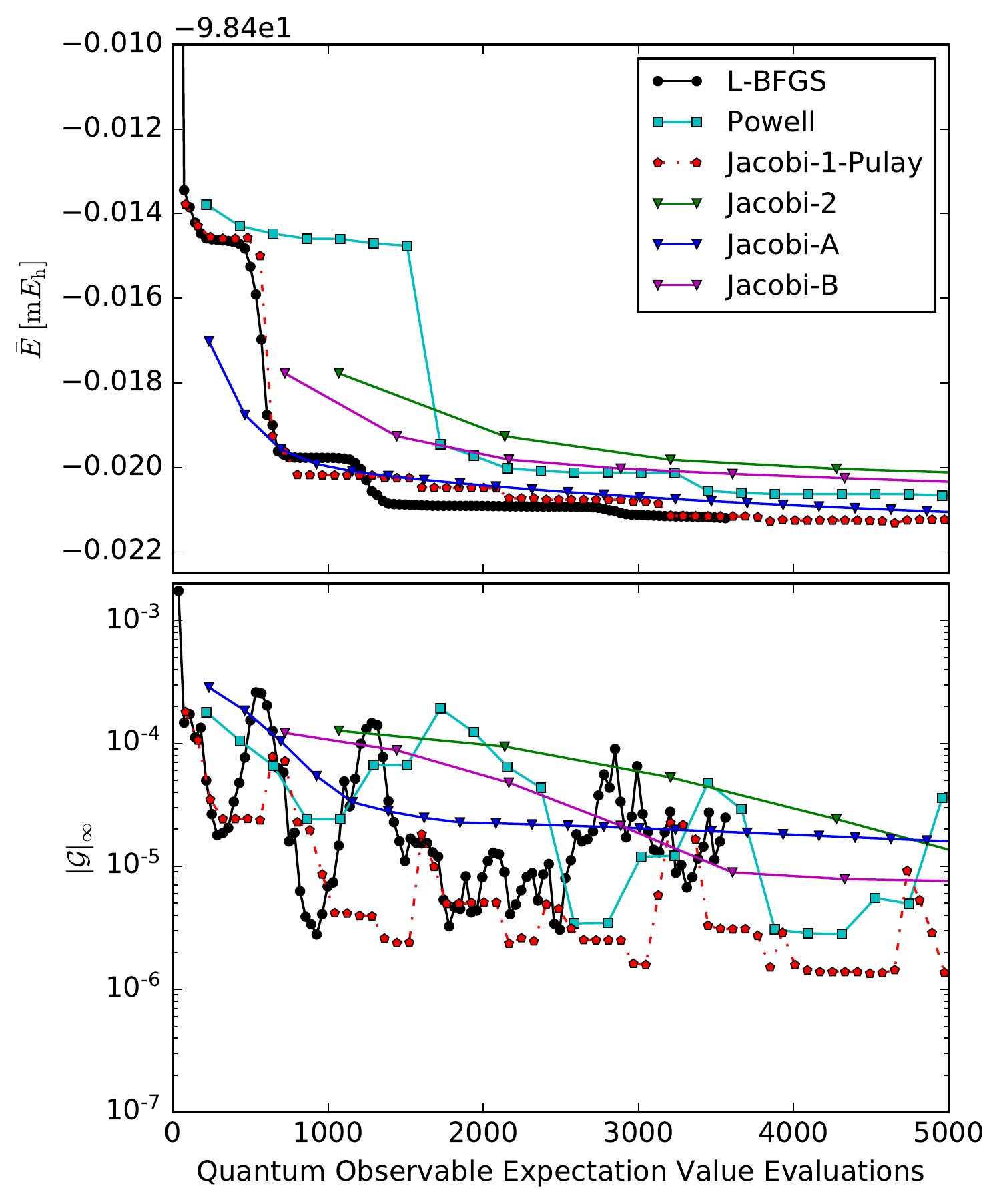}
\caption{State-averaged VQE energy convergence (top) and maximum gradient
element (bottom) as a function of observable expectation value count for the
$N_{\mathrm{state}}=3$ ``hard'' test case. Jacobi-1-Pulay, Jacobi-2, Jacobi-A, and Jacobi-B
methods compared to L-BFGS and Powell.}
\label{fig:final-3B2}
\end{center}
\end{figure}

In Figure \ref{fig:final-3B2}, we compare the early convergence history of the
selected Jacobi methods with L-BFGS and Powell for the ``hard'' test case. Here,
the Jacobi-1-Pulay method retains the prefactor gain over Powell, but the
convergence profile closely resembles L-BFGS. We again point out that we could
improve matters by removing the gradient computation overhead in Jacobi-1-Pulay
by switching to Anderson, if the DIIS flush procedure of Anderson could be
stabilized. We also see the same substantial early-iteration gains for Jacobi-A
as seen for the easy test case. 

Overall, we observe that Jacobi-1-Pulay has an early convergence history that is
several times faster than Powell and never worse than L-BFGS, and we observe
that Jacobi-A provides the lowest-cost route to a robust starting point.

\section{Summary and Outlook}

We have introduced and numerically explored a general class of methods for
variational quantum circuit optimization based on Jacobi-diagonalization-type
sweeps over fully converged sub-optimizations of localized clusters of circuit
parameters. Evaluation of the observable expectation value at small Fourier
quadrature grids allows for the tomography of the observable to be fit to a
simple analytical form, allowing for offline classical optimization over the
free parameters in each cluster. Various cluster strategies are proposed,
including all single angles (Jacobi-1), all pairs of angles (Jacobi-2), and
qubit-locality-based restrictions thereof (Jacobi-A and Jacobi-B). We also
investigated the augmentation of the Jacobi method with Anderson acceleration
through either the traditional Anderson recipe or the Pulay DIIS recipe. We
numerically tested the Jacobi+Anderson methods against the closely-related
gradient-free Powell method and the gradient-based L-BFGS method for a pair of
test cases taken from our MC-VQE+AIEM methodology. Jacobi-1-Pulay was found to
provide an early convergence history that is identical to but several times
faster than Powell, with regard to the number of observable expectation values
required. The early convergence history of Jacobi-1-Pulay is competitive with
and sometimes faster than L-BFGS. Jacobi-A was found to provide an initial
iteration with remarkably low error at correspondingly low cost in terms of
number of observable expectation values required, and may prove highly useful as
a robust starting point that avoids many higher-lying local minima. 

A number of outstanding technical developments remain to be done that might
serve to make the methodology even more useful. Heuristics or guidelines for the
selection of a more-optimal Jacobi sweep pivot order might provide enhanced
convergence rates within each class of Jacobi method. A thorough exploration of
the sieving of insignificant Jacobi-2 pairs to produce alternatives to Jacobi-A
or Jacobi-B might produce additional approaches with costs similar to Jacobi-1
but convergence rates closer to Jacobi-2. It is also worth considering the
addition of certain key 3- and 4-parameter clusters, which are more expensive to
individually optimize, but might drastically improve global convergence. With
regard to the Anderson and Pulay DIIS convergence acceleration explored here, it
is worth considering other pairings of state and error vectors. It is also
necessary to pursue extensions such as EDIIS\cite{kudin2002black} that penalize
non-monotonic convergence behavior and to stabilize the DIIS flush procedure. 

Further afield, it appears that the methodology introduced here is likely just
one of many possible developments in optimization techniques that could make
progress by consideration of the unique mathematical form and constraints of the
optimization problem in variational quantum circuit algorithms. For instance,
the constraint that the tomography is easier to resolve along individual circuit
angles than in arbitrary search directions was the main motivation for the
development of Jacobi-1-Anderson and Jacobi-1-Pulay.  It is plausible that
similar knowledge could be used to lower the cost of the linesearch in Powell's
method. In another direction, one could also consider the use of
non-quadrature-based tomography fitting within the Jacobi steps, which might
prove to be more resilient to device noise channels. In any case, the
introduction of Jacobi fixed-point iteration and Anderson/Pulay DIIS
acceleration techniques provides a promising new direction to explore to meet
the challenge of optimizing variational quantum circuits on NISQ-era quantum
hardware.

\textbf{Note Added in Proof:} While finalizing our manuscript, we became aware
of similar work by Nakanishi, Fujii, and Todo, which was posted on the arXiv
very recently.\cite{Nakanishi:2019:Jacobi} The technical derivations in their
manuscript cover many of the same topics and findings as we do, and Nakanishi et
al. numerically demonstrate an optimization method that appears to be very
similar to one of the methods we describe, Jacobi-1.

\textbf{Acknowledgements:} R.M.P. thanks Dr.~Edward G. Hohenstein, Prof.~Todd J.
Mart\'inez, and Prof.~C. David Sherrill for years of discussions on Jacobi- and
DIIS-type methods for classical electronic structure theory.

\textbf{Conflict of Interest Disclosure:} R.M.P., J.T.I., A.O., and P.L.M. hold
stock/options in QC Ware Corporation.

\bibliography{jrncodes.bib,refs.bib}
\bibliographystyle{aip}

\newpage


% ***** Figures and Tables *****

\end{document}